\documentclass[useAMS,usenatbib]{mn2e}
\usepackage{epsfig,lscape} 
\usepackage{natbib}
\usepackage{amsmath}
\usepackage{graphicx}
\usepackage{color}

\newcommand{\ha}{\hbox{H$\alpha$}}
\newcommand{\hb}{\hbox{H$\beta$}}
\newcommand{\oii}{\hbox{[O\,{\sc ii}]}}
\newcommand{\oiii}{\hbox{[O\,{\sc iii}]}}

\newcommand{\nii}{\hbox{[N\,{\sc ii}]}}

\title[SDSS-IV MaNGA: metallicity gradient in SFGs]{SDSS-IV MaNGA: Modelling the metallicity gradients of gas and stars - radially dependent metal outflow vs IMF }

\author[J. Lian et al.]
{Jianhui Lian$^{1}$\thanks{jianhui.lian@port.ac.uk},
	Daniel Thomas$^{1}$, Claudia Maraston$^{1}$, Daniel Goddard$^{1}$,
	Taniya Parikh$^{1}$, 
	\newauthor J. G. Fern\'andez-Trincado$^{2,3}$, Alexandre Roman-Lopes$^{4}$, Yu Rong$^{5}$, Baitian Tang$^{2}$,
	\newauthor Renbin Yan$^{6}$\\
	$^{1}$Institute of Cosmology and Gravitation, University of Portsmouth, Burnaby Road, Portsmouth, PO1 3FX, UK\\
	$^{2}$Departamento de Astronom\'\i a, Casilla 160-C, Universidad de Concepci\'on, Concepci\'on, Chile \\
	$^{3}$Institut Utinam, CNRS UMR6213, Univ. Bourgogne Franche-Comt\'e, OSU THETA , Observatoire de Besan\c{c}on, BP 1615, 25010 \\
	Besan\c{c}on Cedex, France \\
	$^{4}$Universidad de La Serena, Departamento de Física, Facultad de Ciencias, Cisternas 1200, La Serena, Chile\\
	$^{5}$Chinese Academy of Sciences,National Astronomical Observatories of China, 20A Datun Road, Chaoyang District, Beijing 100012, China\\
	$^{6}$Department of Physics and Astronomy, University of Kentucky, 505 Rose Street, Lexington, KY, 40506-0055, USA
}

\begin{document}
	
	\maketitle
	
	\begin{abstract}
		In our previous work, we found that only two scenarios are capable of reproducing the observed integrated mass-metallicity relations for the gas and stellar components 
		of local star-forming galaxies simultaneously. One scenario invokes a time-dependent metal outflow loading factor with stronger outflows at early times. The other scenario uses a time-dependent IMF slope with a steeper IMF at early times.
		In this work, we extend our study to investigate the radial profile of gas and stellar metallicity in local star-forming galaxies using spatially resolved spectroscopic data from the SDSS-IV MaNGA survey. 
		We find that most galaxies show negative gradients in both gas and stellar metallicity with steeper gradients in stellar metallicity. 
		The {\em stellar} metallicity gradient tend to be mass dependent with steeper gradients in more massive galaxies while no clear mass dependence is found for the {\em gas} metallicity gradient.
		Then we compare the observations with the predictions from a chemical evolution model of the radial profiles of gas and stellar metallicities.
		We confirm that the two scenarios proposed in our previous work are also required to explain the metallicity gradients. Based on these two scenarios we successfully reproduce the radial profiles of gas metallicity, stellar metallicity, stellar mass surface density, and star formation rate (SFR) surface density simultaneously. The origin of the negative gradient in {\em stellar} metallicity turns out to be driven by either radially dependent metal outflow or IMF slope.  
		In contrast, the radial dependence of the {\em gas} metallicity is less constrained because of the degeneracy in model parameters. 
	\end{abstract}
	
	
	\section{Introduction}
	
	Metallicity is a key parameter in galaxy evolution, which plays an important role in many fundamental physical processes that drive the evolution of galaxies, such as gas cooling, star formation, stellar evolution, and dust formation. 
	A correlation of gas metallicity with another fundamental parameter, stellar mass, is known since the 1970s \citep{lequeux1979} and is well-established after several decades \citep{tremonti2004,kewley2008}. 
	More recently, a third dimensional relation between gas metallicity, stellar mass, and star formation rate (SFR) of local star-forming galaxies has also been proposed \citep{ellison2008,mannucci2010,lopez2010,andrews2013}, in which galaxies with higher SFR are typically less metal-enriched at a given stellar mass. Similar to the mass-metallicity relation for the ionized gas (MZR$_{\rm gas}$) in galaxies, people also found that more massive galaxies tend to have higher metallicity in stars, i.e. positive mass-metallicity relation for the stellar component (MZR$_{\rm star}$, \citealt{gallazzi2005,panter2008,thomas2010}). 
	
	In addition to the integrated chemical properties, the spatial distribution, i.e.\ the radial gradient in metallicity is also important to understand the formation history of galaxies. 
	Studying metallicity gradients does not only shed light on the origin of the integrated metallicity scaling relations but also helps to understand how the various internal parts evolve to shape the whole galaxy. Based on long-slit spectroscopy, a negative metallicity gradient (i.e.\ lower metallicity at larger radii) in ionised gas in the interstellar medium (ISM) has been found to be prevalent in local star-forming galaxies \citep{vila-costas1992,zaritsky1994,vanzee1998,moustakas2010}. 
	Likewise, many studies, mostly focusing on early-type galaxies, also found negative gradients in stellar metallicity \citep{carollo1993,davies1993,jorgensen1999,mehlert2003,sanchez2007}.
	
	Integral field unit (IFU) spectroscopy provides a better spatial sampling of a galaxy that enables a more accurate measurement of the gradient in galaxy properties. Early IFU surveys, such as SAURON and ATLAS-3D \citep{bacon2001,cappellari2011} focused on early-type galaxies while many recent and on-going IFU surveys, CALIFA, \citep{sanchez2012}, SAMI \citep{croom2012} and MaNGA \citep{bundy2015} also target star-forming galaxies with {a large galaxy sample}. 
	Based on the CALIFA survey, many studies found a characteristic negative gas oxygen abundance gradient in nearby star-forming galaxies \citep{sanchez2014,ho2015,perez2016,sanchez2016}, which is independent of the galaxy stellar mass. 
	Using what is currently the largest galaxy sample observed by the MaNGA survey, \citet{belfiore2017} revisited the oxygen abundance gradient of local galaxies and found the gradient steepens with stellar mass at the low mass end and flattens at $M_*>10^{10.5}{\rm M_{\odot}}$. 
	As for the stellar metallicity, a negative gradient is usually found based on IFU observations which is consistent with the previous long-slit spectroscopy studies \citep{kuntschner2010,gonzalez2015,goddard2017b,zheng2017}.
	In terms of the mass dependence, using IFU data from the MaNGA survey, \citet{goddard2017b} found that the stellar metallicity gradient in local star-forming galaxies steepens considerably with increasing galaxy mass. 
	
	To explain the observed metallicity gradients in gas and stars, 
	a number of galactic chemical evolution models have been proposed \citep{chiappini2001,fu2009,schonrich2017}, including cosmological chemo-dynamical simulations \citep{rahimi2011,gibson2013,taylor2015,taylor2017}. To reproduce the gas metallicity gradient in local star-forming galaxies, models typically assume a radially dependent gas accretion, no radial matter exchange, and no galactic metal outflow \citep{chiappini2001,fu2009}. 
	As for the stellar metallicity gradient, most theoretical studies focus on explaining the observations in the Milky Way \citep{rahimi2011,gibson2013} or early-type galaxies in the low redshift extragalactic universe \citep{pipino2010}. Only a few theoretical studies attempt to explain both the extragalactic stellar and gas metallicity gradients of star-forming galaxies.
	
	In modelling the chemical evolution of galaxies, it is of great importance to combine the gas and stellar metallicities to constrain the chemical enrichment processes at different epochs.  
	In \citet[][hereafter Paper~I]{lian2017} we present a numerical chemical evolution model that successfully reproduces the observed integrated mass-metallicity relations (MZR) for both gas (MZR$_{\rm gas}$) and stars (MZR$_{\rm stars}$), as well as the observed mass-SFR relation of local star-forming galaxies. Our model is capable of producing a much lower stellar metallicity than gas metallicity and a steeper MZR$_{\rm stars}$ compared to the MZR$_{\rm gas}$. It turns out that only two scenarios can reproduce these metallicity observations simultaneously. One invokes a time-dependent metal outflow loading factor with stronger outflow at early epochs. The alternative scenario requires a time-dependent IMF slope, in which a steeper slope in lower-mass galaxies is needed at early times. In this paper, we extend this work and investigate the spatially-resolved gas and stellar metallicity observations using MaNGA IFU data. In particular, we aim to investigate the chemical evolution history of galaxies as a function of radius by reproducing the radial profiles of gas and stellar metallicity.
	
	The paper is organised as follows. In \textsection 2 we illustrate the sample selection from the MaNGA survey and the derivation of gas and stellar metallicity. We also make a direct comparison between the two metallicity radial profiles in this section. 
	Then we provide a brief introduction to the chemical evolution model in \textsection 3 and show the two effective scenarios that match the observations in \textsection 4. Finally we present our conclusions in \textsection 5.
	Throughout this paper, we adopt the cosmological parameters, $H_0=70\, {\rm km s^{-1} Mpc}^{-1}$, $\Omega_{\Lambda}=0.70$ and $\Omega_{\rm m}=0.30$.
	
	\section{Observations}
	\subsection{Sample selection}
	The MaNGA survey \citep{bundy2015,drory2015,law2015,yan2016a,yan2016b,wake2017} is part of the fourth generation of SDSS surveys \citep{blanton2017} and aims to obtain IFU observation for a representative sample of $10,000$ nearby galaxies with redshift range $0.01<z<0.15$. 
	The MaNGA target sample is selected from the NASA-Sloan Atlas catalogue (NSA, \citealt{blanton2011}) which covers a wide range of galaxy properties \citep{yan2016b,wake2017} and incorporates UV observation from the {\sl GALEX} survey \citep{martin2005}. 
	The observed spectra were reduced using the MaNGA data-reduction-pipeline (DRP, \citealt{law2016}) and then further analysed using the MaNGA data analysis pipeline (DAP, \citealt{westfall2017}).
	
	In this work, we select our star-forming galaxy sample from the latest MaNGA data release 14 (DR14, \citealt{abolfathi2017}) in two steps. All of the integrated galaxy properties used for sample selection are taken from the NSA catalog. {The sample selection of MaNGA galaxies is discussed in detail in \citet{yan2016b}.}
	Firstly, to select a parent galaxy sample, we apply two basic criteria:
	\begin{itemize}
		\item Blue colour with ${\rm NUV}-u<1.4$
		\item Non edge-on system with axis ratio $b/a>0.5$.
	\end{itemize}
	The colour cut is set to select star-forming galaxies according to the the galaxy distribution in ${\rm NUV}-u$ colour space in \citet{lian2016}. 
	The number density distribution of galaxies in ${\rm NUV}-u$ colour shows dramatic number density drops which can be used to roughly classify galaxies into star-forming, green valley, and quiescent populations. These colour cut criteria for star-forming galaxies are consistent with the specific SFR criteria used in Paper I. 
	
	Figure~1 shows the distribution of galaxies in the mass-${\rm NUV}-u$ diagram. The black points represent the entire galaxy sample in MaNGA DR14 with UV detections. Blue points are galaxies selected by the ${\rm NUV}-u$ colour criteria. The contour in the background indicates the distribution of nearby SDSS galaxies with $z<0.05$. By definition, the galaxies selected by the ${\rm NUV}-u$ colour cut are located in the blue star-forming region.  
	The non-edge-on criterion is set to minimise the effects of inclination and dust extinction.
	These criteria lead to 773 star-forming galaxies from the sample of 2744 galaxies observed in MaNGA DR14. 
	
	\begin{figure}
		\centering
        \includegraphics[width=\linewidth]{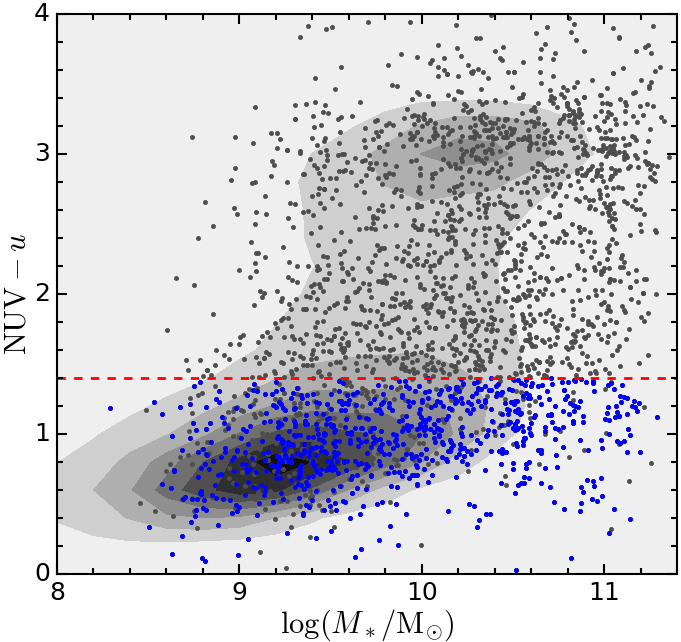}
		\caption{Galaxy distribution in the mass versus ${\rm NUV}-u$ colour diagram. Black dots represent the galaxy sample released with the MaNGA DR14 catalogue. Blue dots are galaxies selected by the ${\rm NUV}-u<1.4$ criterion as indicated by the horizontal dashed line. The underlying contours indicate the distribution of nearby SDSS galaxies with $z<0.05$.  
		}
		\label{figure1}
	\end{figure}
	
	To increase the SNR of spectra, especially for the spaxels in the outskirts, the original low SNR spaxels are binned using a Voronoi binning algorithm \citep{cappellari2003} to reach a minimum ${\rm SNR}=5$ in the $r$-band. 
	To obtain a reliable distribution of galaxy properties, we only use the cells that have a signal-to-noise ratio (SNR) in the strong emission lines (\oii$\lambda$3727,\hb,\oiii$\lambda\lambda$4959,5007,\ha,\nii$\lambda$6584) above 5 and an SNR in the $r$-band stellar continuum above 10. 
		
	After Voronoi binning, there are still cells left that do not satisfy the SNR criteria outlined above. To derive the radial profiles of galaxy properties, we split each galaxy into eight elliptical rings from the center to $1.6r_{\rm e}$ (effective radius) with a width of $0.2r_{\rm e}$. The position angle and ellipticity are taken from the NSA catalogue.  
	{We also require the galaxy to have at least one Voronoi cell that} 
	satisfies the SNR criteria and star-forming classification based on the {demarcation line proposed by \citet{kauffmann2003} in the} BPT diagram \citep{baldwin1981}. 
	This leaves us with {687} galaxies as our final galaxy sample. 
	{We use the Gaussian integral emission line fluxes from the MaNGA DAP and} correct the emission line fluxes for the galactic internal extinction using the Balmer decrement and adopting a Milky Way extinction law \citep{ccm89}.
	
	Figure 2 shows the distribution of all Voronoi cells in the ${\rm log(\oiii\lambda5007/\hb)}$ versus ${\rm log(\nii\lambda6584/\ha)}$ diagram. The galaxy sample is divided into four mass bins {from $10^9$ to $10^{11}\; {\rm M_{\odot}}$ with width of 0.5 dex} and shown in four separate panels. {The number of galaxies per bin are 258, 184, 156, and 89, from the low to the high mass bin, respectively.} In each panel, each data point indicates one Voronoi cell within the corresponding mass bin.  
	The red solid and blue dashed lines represent the demarcation boundaries between star-forming and AGN objects derived by \citet{kewley2001} and \citet{kauffmann2003}. It is interesting to note that there are relatively more regions classified as non-star-forming in more massive galaxies, which is likely a manifest of the downsizing scenario where more massive galaxies tend to have their bulk of star formation at earlier times \citep[e.g.,][]{thomas2005}.
	
	\begin{figure*}
		\centering
		\includegraphics[width=13cm]{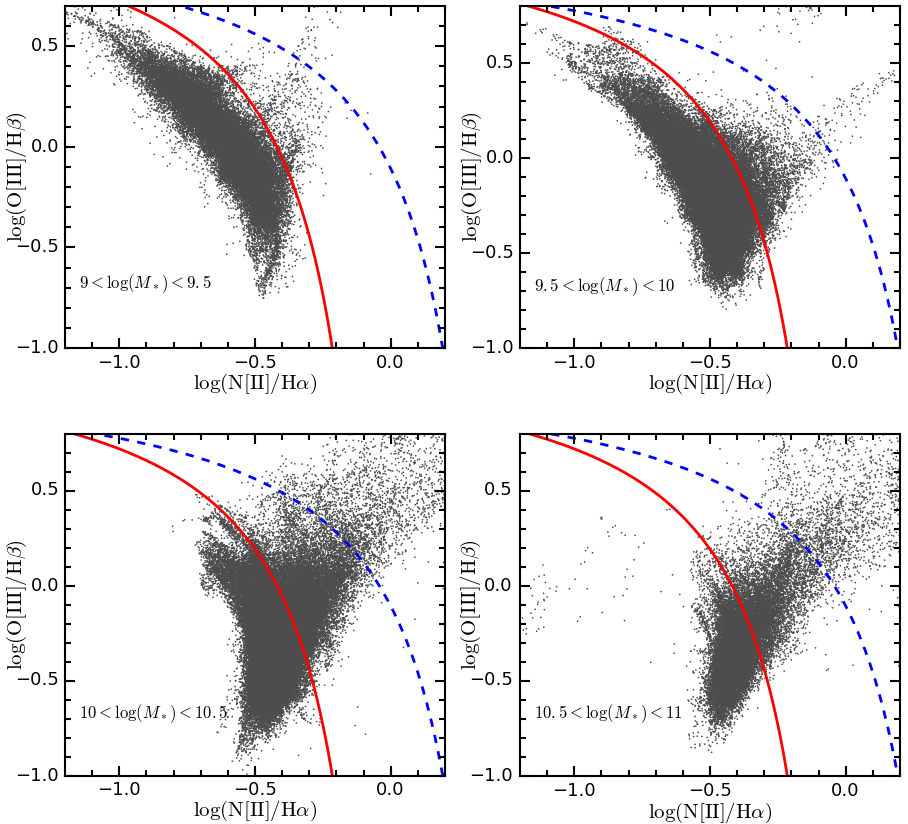}
		\caption{Galaxy distribution in the BPT diagram. The galaxy sample is divided into four mass bins and shown in four panels, separately. In each panel, each point indicates one Voronoi cell within the corresponding mass bin. The red solid and blue dashed lines represent the demarcation boundaries between the star-forming and AGN objects derived by \citet{kauffmann2003} and \citet{kewley2001}, respectively.  
		}
		\label{figure2}
	\end{figure*}
	
	\subsection{Metallicity determination}
	\subsubsection{Gas metallicity}
	There are many methods proposed to measure the gas metallicity of emission-line galaxies \citep{kewley2008,maiolino2008}.
	One of the most reliable methods is the so-called `Te' method which utilises the anti-correlation between electron temperature
	and metal abundance. To obtain the electron temperature, two excitation lines of the same ion are needed. 
	The most widely-used pair of excitation lines are \oiii$\lambda$4363 and \oiii$\lambda$5007. 
	However, since the \oiii$\lambda$4363 line is weak, many other empirical methods calibrated to the Te method are proposed.
	
	These empirical methods use a single or a combination of strong emission line ratios for the calibration.
	Some of the most-widely used are the `N2 method' (\nii$\lambda$6584/\ha; \citealt{pettini2004}), the `O3N2 method' ((\oiii$\lambda$5007/\hb)/(\nii$\lambda6584$/\ha); 
	\citealt{pettini2004}), and the `R23 method' ((\oii$\lambda$3727+\oiii$\lambda\lambda$4959,5007)/\hb; \citealt{pilyugin2005}).
	Rather than calibrating emission line ratios to the metallicity determined by Te method, some studies use similar 
	emission line ratios but calibrate to theoretical photo-ionisation models (e.g., \citealt{m91,kewley2002}). 
	
	However, a large discrepancy by up to 0.7 dex exists between these calibrations \citep{shi2005,kewley2008}. 
	Generally, the theoretical methods lead to much higher metallicities than the empirical methods. 
	The physical origin of this discrepancy is still not fully understood. 
	Following our study in Paper I,
	to ensure our result is independent of this discrepancy, we adopt two methods for the measurement of gas metallicity, namely the 
	theoretical R23 method from \citet{kewley2002} and the empirical `N2' method calibrated by \citet{pettini2004}. 

	\subsubsection{Stellar metallicity}
	
	The stellar metallicity is an important observable as it carries the information of the early chemical properties of galaxies. 
	However, the stellar metallicity of a galaxy is difficult to measure due to well known degeneracies between age, metallicity and dust when analysing galaxy spectra
	{and broadband galaxy photometry}. Yet, modern advancements in stellar population modelling and improvements in the quality of observational data have enabled significant advances in the analysis of stellar population parameters, allowing robust measurements of metallicity to be obtained.
	In this work, we adopt the stellar metallicities from \citet{goddard2017b} derived from using the full spectral fitting code FIREFLY \citep{wilkinson2015,wilkinson2017} and the stellar population models of Maraston and Str\"omb\"ack (\citealt{maraston2011}, M11) with a Kroupa initial mass function \citep[IMF][]{kroupa2001}.
	
	FIREFLY uses a $\chi^2$ minimsation technique to fit combinations of Simple Stellar Population (SSP) models to an input galaxy spectrum. The code uses an iterative algorithm to combine arbitrarily weighted linear combinations of SSPs, in order to find the best fit model given the data and employs minimal priors, allowing maximal exploration of the parameter space \citep[for more detail see][]{goddard2017b,wilkinson2017}. This has been shown to be a good way to accurately recover the properties of galaxies and globular clusters \citep{wilkinson2017}. {For a comparison between FIREFLY and other stellar population synthesis models we refer the reader to \citet{goddard2017b} and \citet{wilkinson2017}.}
	
	\subsection{Gas and stellar metallicity comparison}
	
	\subsubsection{2D maps}
	\begin{figure*}
		\centering
		\includegraphics[width=18cm]{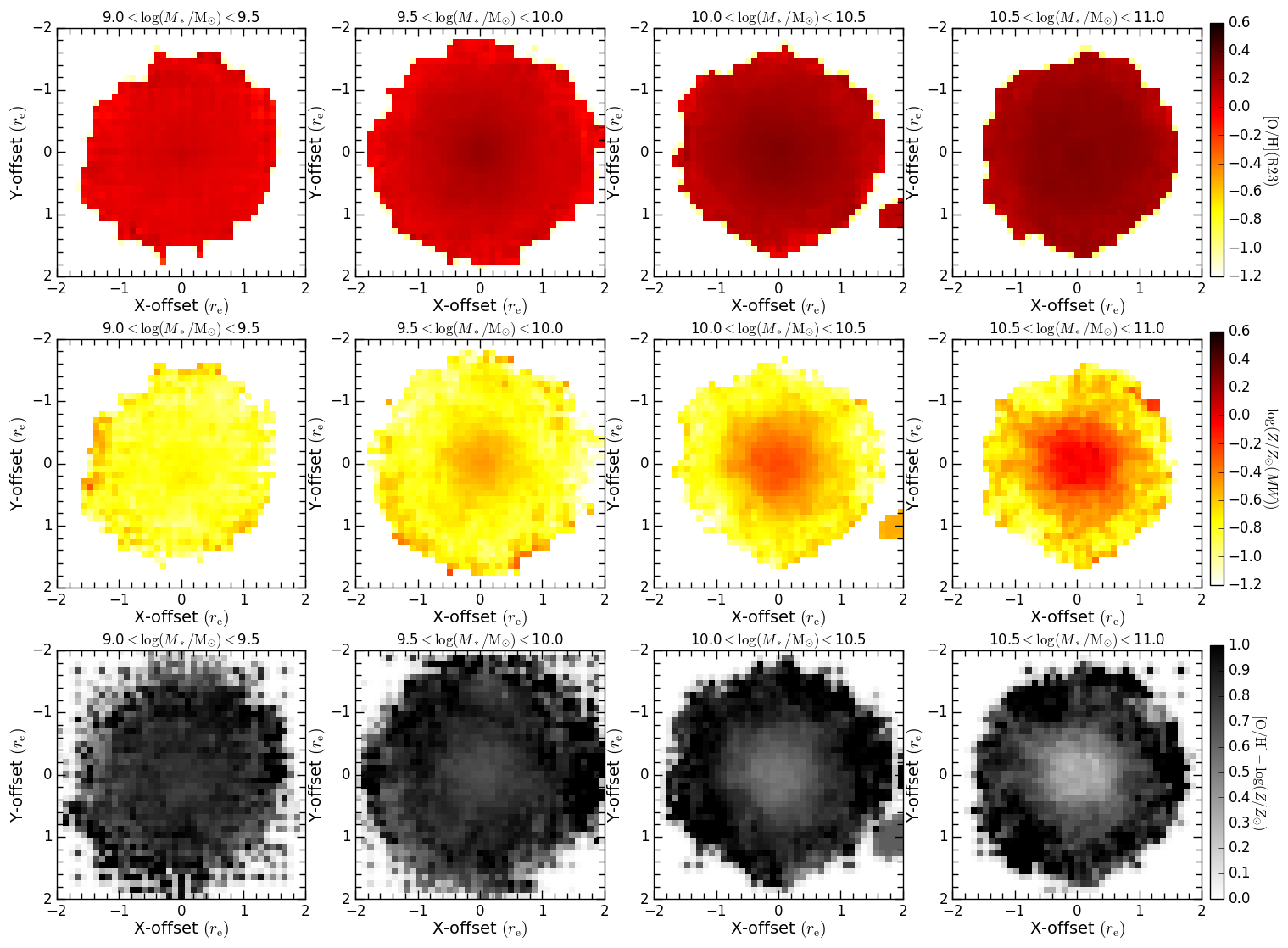}
		\caption{Stacked 2D maps of gas (top row) and stellar metallicity (middle row) and their difference (bottom row) as a function of galaxy stellar mass. Each column indicates one mass bin from low mass at the left-hand side to high mass at the right-hand side. The coordinates of the maps are in units of effective radius. The gas metallicity is derived using the theoretical R23 method.}
		\label{figure3}
	\end{figure*}
	
	Before modelling in detail the radial metallicity profiles, it is worth showing the 2D maps of gas and stellar metallicities to visualise the spatial distribution. For this purpose of visualisation, we stack the metallicity maps of galaxies in our sample in two steps. The first step is to rescale the original metallicity maps of each galaxy to their effective radius, $r_{\rm e}$. Then we define a new 2D grid and find the Voronoi cells of galaxies that are located in each spaxel of the new grid. We take the median metallicity of these Voronoi cells as the metallicity in each spaxel. {Since the 2D maps are only used for visualization, we do not deproject the galaxies to be exactly face-on before stacking. Position angle and ellipticity are taken into account, though, when producing the 1-dimensional profiles for the scientific analysis.}
	
	Figure 3 shows the resulting 2D stacked maps of gas (top row) and stellar metallicity (middle row) and their difference (bottom row) as a function of galaxy stellar mass (panels from left to right). We split the galaxy sample into {the same} four mass bins as {in} Figure 2.
	Each column indicates one mass bin from low mass at the left-hand side to the high mass at the right-hand side. These maps are smoothed by a Gaussian kernel with a width of half a spaxel to remove 
	pixelised features at very large radii.
	
	It can be seen that the gas metallicity measured through the R23 method is much higher than the stellar metallicity by up to $\sim$0.7 dex (see Paper~I). Figure~3 shows that this difference is radially dependent with a significantly smaller difference in galaxy centres. Regarding the spatial distribution, gas metallicity generally exhibits a relatively shallow negative gradient with no significant dependence on mass. 
	In contrast, for the stellar metallicity, we can see a much steeper gradient which is strongly dependent on stellar mass with a steeper gradient in more massive galaxies. This is consistent with the mass-dependence of the stellar metallicity gradient found by MaNGA for late-type galaxies \citep{goddard2017a,goddard2017b}. It is also interesting to note that the {\em stellar} metallicity of massive star-forming galaxies is distinctively higher in the centre than in the outer disc, while the difference in gas metallicity is small. This suggests that galaxy bulges in massive star-forming galaxies form at early times and have distinctive metal enrichment histories from the discs.   
	
	\subsubsection{1D profiles}
	For the analysis in this work we use 1D radial profiles of gas and stellar metallicity.
	To obtain these, we take the median measurement of the Voronoi cells of galaxies in each mass and radial bin. 
	{We use elliptical radial bins to take the inclination of galaxies into account. The same mass bins are used} for the stacked maps in Figure 3. The median effective radii $r_{\rm e}$ are 2.9, 3.9, 5.5, and 8.5 kpc for galaxies in the four mass bins, respectively. Figure 4 shows a direct comparison of the gas and stellar metallicity radial profiles for galaxies in each mass bin. The shaded region indicates the {$1\sigma$} error of the median value in each mass and radial bin {which is obtained through bootstrapping}. The four different mass bins are indicated by the different colours. The two gas metallicity calibrations, the theoretical R23 and the empirical N2 methods, are shown in the left-hand and right-hand panels, respectively.
	
	\begin{figure*}
		\centering
		\includegraphics[width=\textwidth]{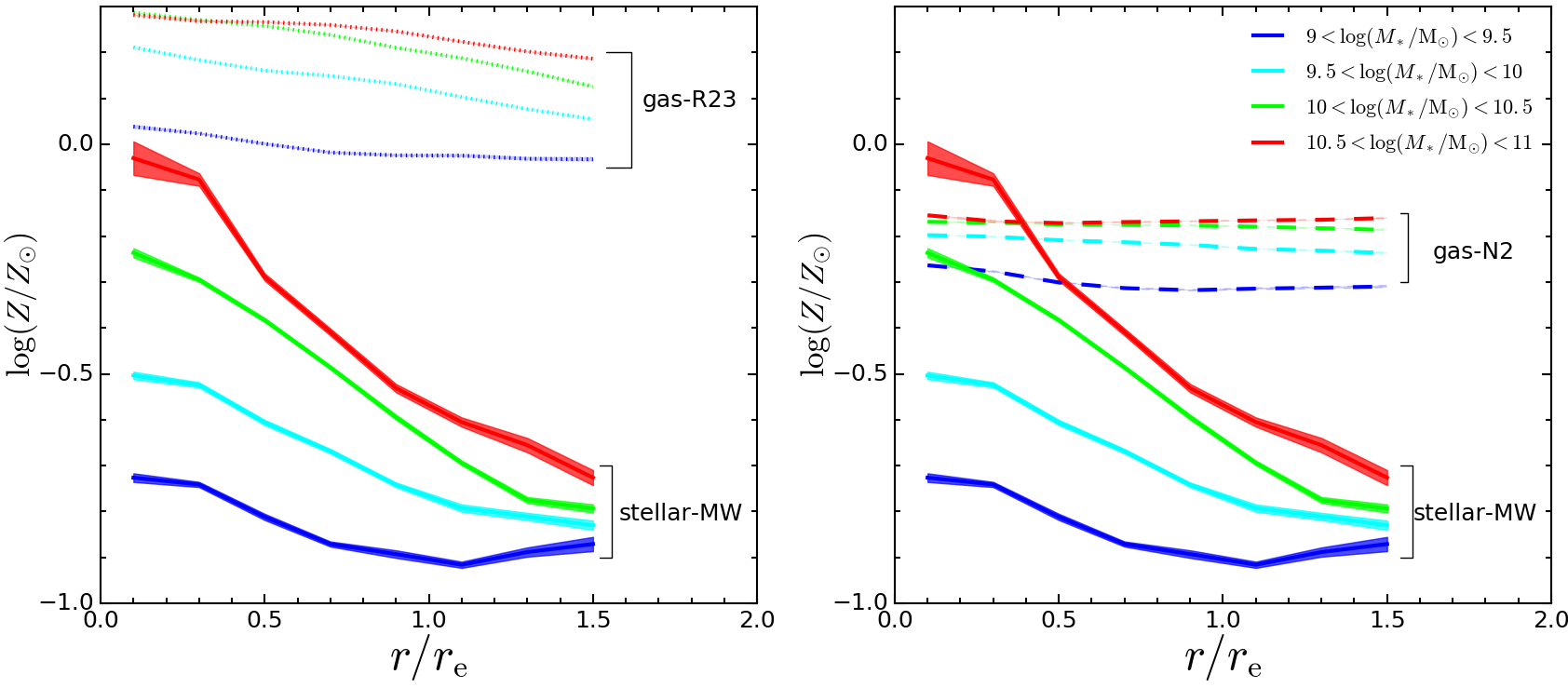}
		\caption{Comparison between the median gas and stellar metallicity radial profiles in four mass bins. Two gas metallicity calibrations, theoretical R23 and empirical N2, are used and shown in two panels, respectively. The mass range of each mass bin is indicated in the legend in the right panel.}   
		\label{figure4}
	\end{figure*}
	
	It can easily be appreciated that gas metallicity is always higher than the stellar metallicity at all radii in agreement with the global properties discussed in Paper~I. As discussed there, this is indeed to be expected if the metal enrichment is a monotonous process and the gas metallicity is set primarily by the instantaneous chemical enrichment rather than by the past history of the system \citep{lilly2013}.
	Although a higher gas metallicity is expected, we show in Paper I that the relatively large difference poses a challenge to chemical evolution models that generally fail to reproduce gas and stellar metallicity simultaneously.
	
	In Paper I we find that this difference between gas and stellar metallicity is dependent on stellar mass with the difference being largest in the lowest-mass galaxies. In Figure 4, this trend can still be appreciated. In addition to this dependence on stellar mass, we can now also see a clear dependence on radius. The discrepancy between gas and stellar metallicity clearly increase with increasing radius. In other words, the stellar metallicity gradient is much steeper than the gas metallicity gradient. A key motivation of the present paper is to reproduce these radial profiles to set constraints on the galaxy chemical enrichment and formation histories.
	
	It is also interesting to note that the stellar metallicity gradient significantly steepens with increasing stellar mass while the gas metallicity gradient hardly changes. This is consistent with the 2D maps shown in Figure 3. This trend also implies that galaxy bulges tend to have significantly enhanced stellar metallicities but only slightly higher gas metallicities compared to the disc, suggesting different metal enrichment histories for different galaxy components. {The mass dependence of the stellar metallicity gradient in early-type galaxies is similar but much milder compared to the late-type galaxies as shown in \citet{goddard2017b}. More complicated picture of the mass dependence in early-type galaxies is proposed by (\citealt{spolaor2010,kuntschner2010}).} Another interesting fact worth noting in Figure 4 is that, unlike the inner region of galaxies which have a stellar metallicity strongly dependent on the global stellar mass, the outer discs beyond 1 $r_{\rm e}$ tend to show similar stellar metallicities without strong dependence on galaxy mass. This suggests a universal formation history of galaxy discs largely independent of galaxy mass.

	\section{Chemical evolution model}
	In Paper~I we constructed a full galactic chemical evolution (GCE) model to reproduce the integrated properties of local star-forming galaxies, including 
	mass, SFR, gas and stellar metallicity. To study the metallicity gradient in the present paper, we modify our chemical evolution model to account for the chemical evolution at different radii. Here, we briefly introduce the basic ingredients of the chemical evolution model. For a more detailed description we refer the reader to \textsection 3.1 in Paper~I. 
	
	\subsection{Model ingredients}
	
	Generally, a numerical chemical evolution model accounts for the main processes that regulate metal enrichment, including gas inflow, gas outflow and mass ejection from stars including AGB stellar winds, Type~II supernovae (SN-II), and Type~I supernovae (SN-I). In our model we adopt a single phase of pristine-gas inflow that declines exponentially with cosmic time. As a result, the mass assembly history and star formation history of the model galaxy are mainly regulated by this gas inflow process. The SFR at each time is calculated based on the gas mass surface density and the Schmidt-Kennicutt (KS) star-formation law \citep{kennicutt1998}. Observed galaxy sizes (effective radii) are used to convert between the surface density and integrated properties.
	
	The gas outflow, which is difficult to model, is parameterised by the fraction of mass ejected from stars and expelled from the galaxy. This metal outflow fraction is linearly proportional to the metal outflow mass loading factor defined in the literature. 
	Regarding the mass ejection by stellar processes, which usually depend on the mass and the metallicity of the progenitor star, 
	we adopt the metal yields of AGB stars from \citet{ventura2013}
	, yields of SN-I from \citet{iwamoto1999}, and mass ejection of SN-II derived by \citet{portinari1998} based on SN nucleosynthesis models of \citet{woosley1995}.
	
	The integrated metal production of a single stellar population depends on the IMF \citep{thomas1999}. In the chemical evolution model we adopt a double-power law IMF and allow the slope at the low-mass end ($M_*<0.5{\rm M_{\odot}}$) and at the high mass end ($M_*>0.5{\rm M_{\odot}}$) to vary. 
	{The low-mass and high-mass cut-offs of the IMF are 0.01 ${\rm M_{\odot}}$ and 120 ${\rm M_{\odot}}$, respectively.}
	There are three basic assumptions in this chemical evolution model. 
	First, as mentioned above, we assume a single phase of pristine gas inflow which dominates the star formation history of the model galaxy. 
	Second, we only consider newly ejected mass from stars to be expelled out of the galaxy by the galactic outflow. Lastly, no radial mass exchange is assumed. 
	
	\subsection{Model parameters}
	So far, we have the following eight free parameters:
	\begin{itemize}
		\item $A_{\rm ks}$, $a_{\rm ks}$: coefficient and power index of the KS star-formation law.
		\item $A_{\rm inf}$, $\tau_{\rm inf}$: initial inflow strength and its declining time scale. \item $f_{\rm out}$: metal outflow fraction.
		\item $\alpha1$, $\alpha2$: IMF slope at the low-mass and high-mass ends, respectively.
		\item $t_{\rm delay}$: mixing time delay of the stellar ejecta with ISM.
	\end{itemize}
	The coefficient of the KS law is normalised to the value of $2.5\times10^{-4} {\rm M_{\odot}yr^{-1}kpc^{-2}}$ as derived by \citet{kennicutt1998}. 
	As we found in Paper~I, a simple chemical evolution model with a KS star-formation law, Kroupa IMF and no outflow is too efficient in metal production at early times and thus results in a stellar metallicities much higher than the observed values. Moreover, the predicted stellar metallicities are strongly coupled with the gas metallicities and therefore the model cannot reproduce the different dynamic ranges of observed gas and stellar metallicity simultaneously. To decouple the gas and stellar metallicity and reproduce the observations, either a time-dependent metal outflow or IMF is needed (see Paper I). 
	
	\subsection{Radial dependence}
	To reproduce the radial profile of gas and stellar metallicity, we use eight GCE models to mimic eight radial bins from the centre to 1.6$r_{\rm e}$ with a bin width of 0.2$r_{\rm e}$. With this ingredient, all of the parameters have the additional freedom of varying as a function of radius. The initial inflow strength will be strictly constrained by the stellar mass in each radial bin derived from the observational data, while the inflow time-scale will be tuned to match the observed SFR.   
	For simplicity, we fix the power index of the KS law $a_{\rm ks}$ to be 1.5 which is close to the value of 1.4 obtained by \citet{kennicutt1998}. 
	The rest of the free parameters will be tuned to match the observed radial profiles in gas and stellar metallicity. 
	
    {The best-fitting model and parameters as a function of mass and radius are chosen by visual inspection. A radial dependence has been tested {\it for all parameters} in the search for the best-fitting model, but, for the sake of simplicity, we only use radially dependent parameters when necessary to reproduce the observations. As a result, many parameters in the best-fitting model will be invariable as a function of radius and will be listed as a constant in the tables. {The form of the radial dependence of the gas inflow rate is assumed to be exponential to match the typically exponential profiles of the surface mass density. The radial dependence of other parameters is chosen to be linear for the sake of simplicity.}
	
	\subsection{Parameter degeneracies}
	In Paper~I we explored the parameter space to understand the potential degeneracy between these model parameters. Generally, $A_{\rm ks}$, $a_{\rm ks}$, and $A_{\rm inf}$ are degenerate since they regulate the metallicity in the same way by affecting the star formation efficiency (SFE). Higher values of these three parameters lead to higher SFE and usually result in higher gas and stellar metallicity. Aside from the SFE, the metallicity (both in gas and stars) could also be enhanced by adopting a lower metal outflow fraction or a flatter IMF slope. Because of the degeneracy between these parameters, three models with various {\sl present} SFE, {\sl present} metal outflow fraction or {\sl present} IMF slope are all capable of reproducing the mass and radial dependence of the gas metallicity. 
	
	\subsection{Failure of simple test models}
	
	\begin{figure*}
		\centering
        \includegraphics[width=17cm]{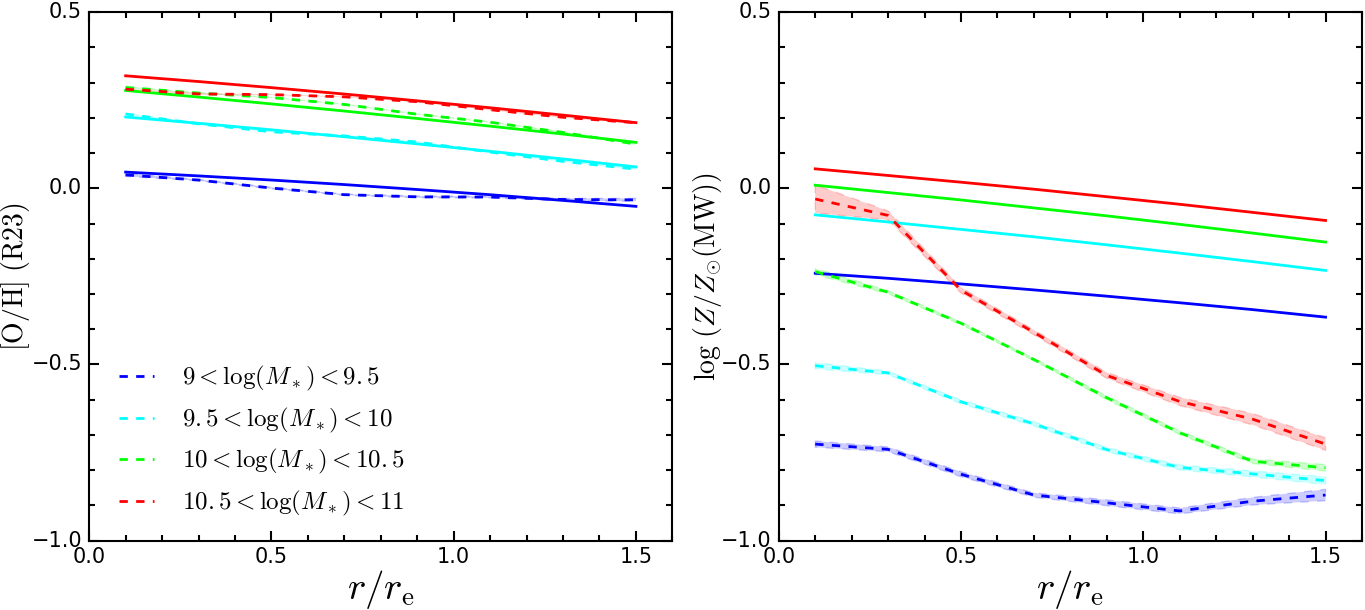}
		\includegraphics[width=17cm]{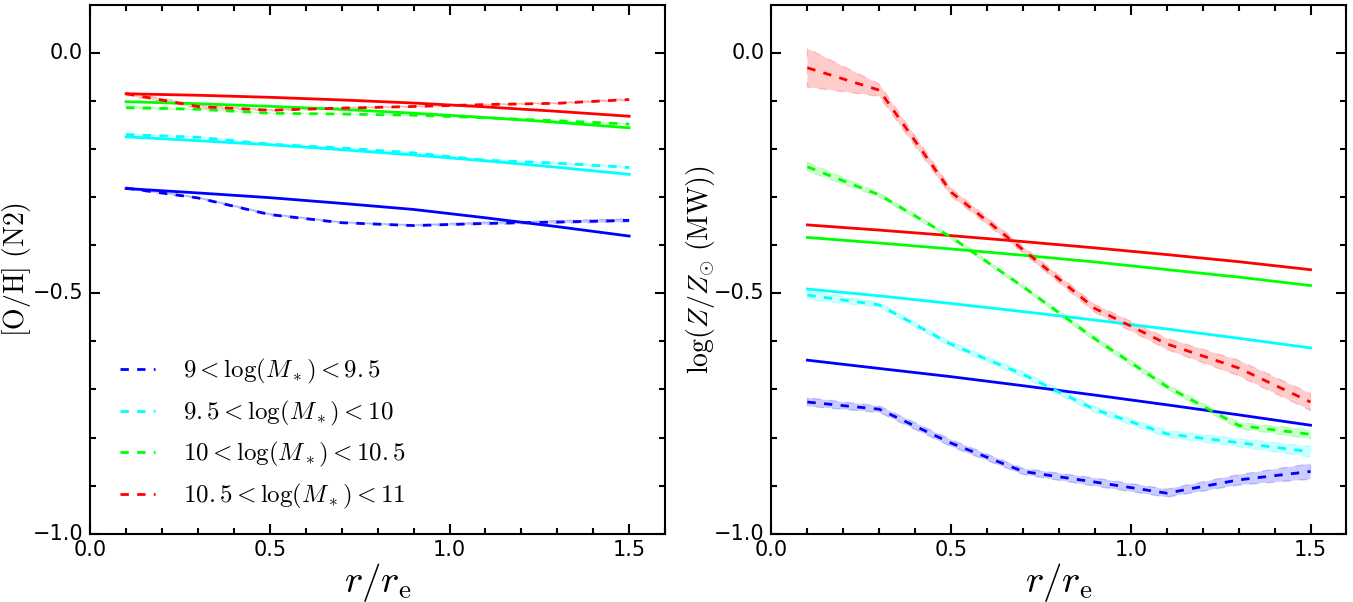}
		\caption{Comparison between the MaNGA data and model predictions for test chemical evolution {models} of the radial profile of gas metallicity and mass-weighted stellar metallicity (right-hand panel). The top and bottom panels show results where gas metallicity has been measured through the theoretical R23 and empirical N2 methods, respectively. Dashed lines represent the observed radial profiles, while solid lines are the model predictions. Different colours represent different mass bins as illustrated by the legend in the bottom left-hand panel. {These test models are tuned to only match the gas metallicity observations.}
		}
		\label{figure5}
	\end{figure*}
	
	Figure 5 presents the comparison between MaNGA data and model prediction for a simple test model of the radial profile of gas and stellar metallicities. The top and bottom panels show results where the gas metallicity has been measured through the theoretical R23 and empirical N2 methods, respectively. 
	The models assume a radially dependent gas accretion and no time-dependence of metal outflow fraction or IMF slope. The model parameters are tuned to reproduce the radial profile of gas metallicity as derived through the R23 method (top panels) or through the N2 method (bottom panels, see also Table 1 and Table 2). {For simplicity, we only show the test model with varying metal outflow fraction when necessary keeping the IMF fixed to Kroupa (for a discussion on the test model with varying IMF we refer the reader to \textsection4 in Paper I).} These test models are close to the chemical evolution model typically used in the literature to reproduce the gas metallicity gradient. The dashed lines indicate the observations while the solid lines represent the predictions of the model. 
	Different colours show the galaxies in different mass bins, from the lowest mass bin in blue at the bottom to the highest mass bin in red at the top. Since the radial profiles of the mass surface density in the four mass bins can be approximated by a power law function, we adopt a surface density of inflow rate that declines exponentially with radius. 
	
	It can be seen that the model matches the radial profile of {\em gas} metallicity very well. The negative gradient in gas metallicity is produced by a negative gradient in gas accretion rate, i.e.\ SFE. However, the model fails to reproduce the {\em stellar} metallicity profiles. Firstly, we can see that the model predicts a stellar metallicity gradient that is much flatter than the observation for both gas metallicity calibrations. When adopting the gas metallicity from the R23 method, the models also overpredict the stellar metallicity by up to 0.5 dex at all radii and for all galaxy masses. This over-prediction problem implies that the model is too efficient in metal {enrichment} at early times which results in relatively high {\em present} stellar metallicity. Using the gas metallicity determined through the N2 method, this over-prediction problem seems to be partially relieved but the model still produces stellar metallicity gradients that are too shallow, and it over-predicts the stellar metallicity in the lowest-mass galaxies and the outer disc. This over-prediction problem is equivalent to the problem in matching the global mass-metallicity relations as discussed in Paper I.

	\begin{table*}
		\caption{Parameter {values adopted in} the test chemical evolution {models} in Figure 5.}
		\label{Table1}
		\centering
		\begin{tabular}{l c c c c c c c c}
			\hline\hline
		[O/H] calibration & mass bin & $A_{\rm ks}$ & $A_{\rm inf}$ $^{a}$ & $\tau_{\rm inf}$ & $f_{\rm out}$ & $\alpha1$ & $\alpha2$ \\ 
		- &	log(${\rm M_{\odot}}$) & - & ${\rm M}_{\odot}{\rm yr^{-1}}$ & Gyr & - & - & - \\
			\hline
		R23 & $[9,9.5]$ & 0.10 & 6$\frac{r}{r_{\rm e}}e^{-\frac{r}{r_{\rm e}}}$ & 5 & 0.52 & 1.3 & 2.3 \\
		R23 &	$[9.5,10]$ & 0.06 & 13$\frac{r}{r_{\rm e}}e^{-\frac{r}{r_{\rm e}}}$ & 5 & 0.19 & 1.3 & 2.3 \\
		R23 &	$[10,10.5]$ & 0.06 & 23$\frac{r}{r_{\rm e}}e^{-\frac{r}{r_{\rm e}}}$ & 5 & 0.0 & 1.3 & 2.3 \\
		R23 &	$[10.5,11]$ & 0.10 & 25$\frac{r}{r_{\rm e}}e^{-\frac{r}{r_{\rm e}}}$ & 5 & 0.0 & 1.3 & 2.3 \\
		\hline
	    N2 &	$[9,9.5]$ & 0.10 & 5$\frac{r}{r_{\rm e}}e^{-\frac{r}{r_{\rm e}}}$ & 5 & 0.78  & 1.3 & 2.3 \\
		N2 & $[9.5,10]$ & 0.10 & 13$\frac{r}{r_{\rm e}}e^{-\frac{r}{r_{\rm e}}}$ & 5 & 0.72 & 1.3 & 2.3 \\
		N2 & $[10,10.5]$ & 0.13 & 26$\frac{r}{r_{\rm e}}e^{-\frac{r}{r_{\rm e}}}$ & 5 & 0.68 & 1.3 & 2.3 \\
		N2 & $[10.5,11]$ & 0.20 & 26$\frac{r}{r_{\rm e}}e^{-\frac{r}{r_{\rm e}}}$ & 5 & 0.67 & 1.3 & 2.3 \\
		\hline	
			\hline
		\end{tabular}\\ 
		Note $^a$: The radial dependence of parameters are illustrated as {an exponential or a linear} function of radius in the table.\\	 	
	\end{table*}
	
	
	\section{Results}
	
	To reproduce the relatively low stellar metallicity, it is necessary to suppress metal enrichment at early epochs without changing the star formation history significantly. In Paper~I we identified two scenarios that are able to achieve that. One is a time-dependent metal outflow loading factor with stronger outflow at early times. The other viable scenario requires a time-dependent IMF slope, such that a steeper IMF slope is assumed at early times. 
	
	\begin{figure}
		\centering
		\includegraphics[width=\columnwidth]{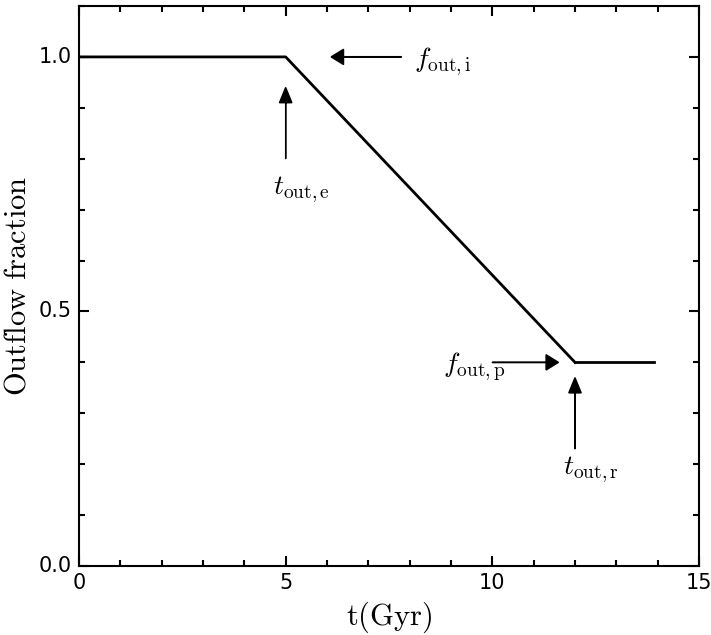}
		\caption{Schematic plot to illustrate the function adopted for the time evolution of the metal outflow fraction. The present metal outflow fraction $f_{\rm out}$ and the transition time $t_{\rm out}$ are marked in the plot. 
		}
		\label{figure6}
	\end{figure}
	
	\subsection{Variable outflow model}
	Following our study in Paper~I, we use a modified linear function to describe the time evolution of the metal outflow fraction as illustrated in Figure 6. Before an early transition {time} $t_{\rm out,e}$, the metal outflow fraction is set to be a high value $f_{\rm out,i}$ (usually 100\%). The gas and stellar metallicity are extremely low during this period. After this transition {time}, the metal outflow fraction begins to decrease linearly to a lower value $f_{\rm out,p}$ until reaching a recent {transition time} $t_{\rm out,r}$ and then staying constant afterwards. The first period with a constant high outflow fraction is needed to efficiently suppress the metal {enrichment} at early times. 
	
	As we discussed above, the dramatically different gradients in gas and stellar metallicity in massive galaxies imply that gas and stellar metallicity are decoupled with different drivers. Since the gas metallicity is generally driven by recent metal enrichment processes, we use the second period with a lower constant outflow fraction to separate the parameters that drive the gas and stellar metallicity as much as possible. 
	
	\begin{figure*}
		\centering
		\includegraphics[width=\textwidth]{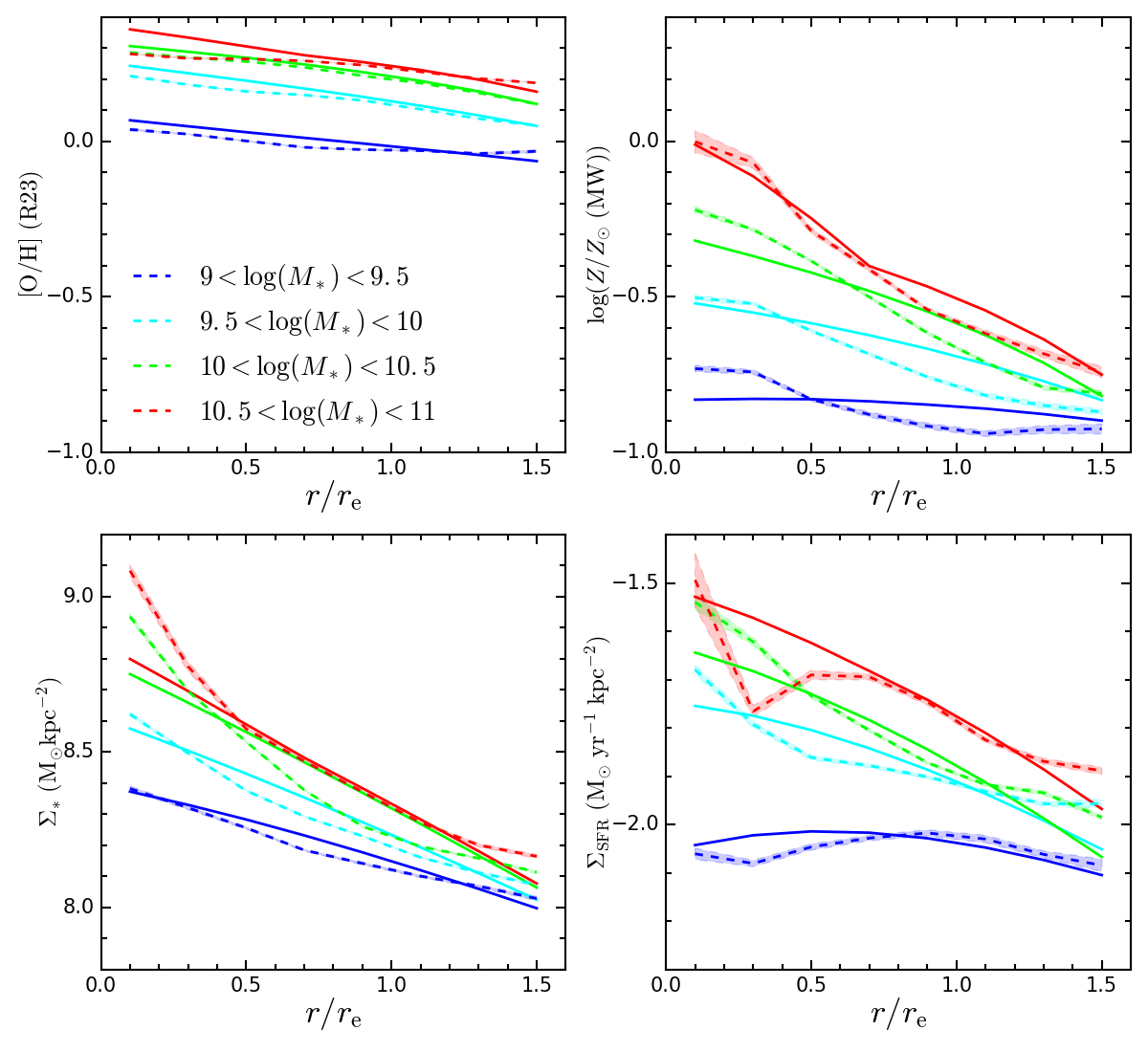}
		\caption{Comparison between MaNGA data and the prediction of models with time-dependent metal outflow fractions. Radial profiles of gas metallicity (from the R23 method), mass-weighted stellar metallicity, mass surface density, and SFR surface density are shown. The observational data are shown as dashed lines while the model predictions are represented by solid lines. Different colours represent different mass bins as indicated in the legend in the top {left-}hand panel. {The parameters adopted in the variable outflow model are listed in Table 2.}
		}
		\label{figure7}
	\end{figure*}
	
	\begin{table*}
		\caption{Parameter value of the time-dependent metal outflow model in Figure 7.}
		\label{Table2}
		\centering
		\begin{tabular}{l c c c c c c c c c c c c}
			\hline\hline
			mass bin & $A_{\rm ks}$ & $A_{\rm inf}$ & $\tau_{\rm inf}$ & $f_{\rm out,i}$ $^a$ & $f_{\rm out,p}$ & $t_{\rm out,e}$ & $t_{\rm out,r}$ & $\alpha1$ & $\alpha1_{\rm i}$ & $\alpha2$&$\alpha2_{\rm i}$ \\ 
			log(${\rm M_{\odot}}$) & - & ${\rm M}_{\odot}{\rm yr^{-1}}$ & Gyr & - & - & Gyr & Gyr &- & - & - & - \\
			\hline
			$[9,9.5]$ & 0.45 & 2.2$\frac{r}{r_{\rm e}}e^{-0.9\frac{r}{r_{\rm e}}}$ & 3.3+1.7$\frac{r}{r_{\rm e}}$ & 1 & 0.28 & 4.5+1.3$\frac{r}{r_{\rm e}}$ & 13.7 & 1.3 & 1.3 & 2.3 & 2.3 \\
			$[9.5,10]$ & 0.50 & 4$\frac{r}{r_{\rm e}}e^{-1.1\frac{r}{r_{\rm e}}}$ & 3.8+1.5$\frac{r}{r_{\rm e}}$ & 1 & 0  & 3.5+2.6$\frac{r}{r_{\rm e}}$ & 13.7 & 1.3 & 1.3 & 2.3 & 2.3 \\
			$[10,10.5]$ & 1 & 11$\frac{r}{r_{\rm e}}e^{-1.3\frac{r}{r_{\rm e}}}$ & 4.0+1.5$\frac{r}{r_{\rm e}}$ & 1 & 0 &  1.1+4.5$\frac{r}{r_{\rm e}}$ & 13.7 & 1.3 & 1.3 & 2.3 & 2.3 \\
			$[10.5,11]$ & 1 & 24$\frac{r}{r_{\rm e}}e^{-1.3\frac{r}{r_{\rm e}}}$ & 4.4+2.0$\frac{r}{r_{\rm e}}$ & 0.6+0.6$\frac{r}{r_{\rm e}}$ & 0 & 0.0+5$\frac{r}{r_{\rm e}}$ & 12.7 & 1.3 & 1.3 & 2.3 & 2.3 \\
			\hline
		\end{tabular}\\
		Note $^a$: {Metal} outflow fraction that exceeds 100\% will be set to 100\%.\\	
	\end{table*}
	
	\subsubsection{Radially dependent in- and outflow}
	Figure 7 shows the comparison between the observations and predictions of the GCE model with a time-dependent metal outflow fraction for 
	the radial profiles of gas (from the R23 method) and mass-weighted stellar metallicity, stellar mass surface density, and SFR surface density.  
	The dashed lines indicate the observational data while the solid lines represent the predictions of the model.  
	As listed in {Table 2}, the model parameters are finely tuned to match the radial profiles of all four observables, gas and stellar metallicities as well as stellar mass surface density and SFR surface density.
	
	To reproduce the surface mass density profile, we simply assume that the inflow rate declines exponentially with radius {($A_{\rm inf}(r) \propto e^{-r/r_{\rm e}}$)}. 
	From the bottom left-hand panel of Figure 7 we can see that more massive galaxies tend to have steeper radial profiles in surface mass density, suggesting they have steeper gradients in the gas inflow rate.
	It is also interesting to note that we need a radially dependent inflow time scale with longer time scales at larger radii in order to match the radial profiles in SFR (which is flatter than the surface mass density). This is consistent with the inside-out growth scenario proposed by \citet{lian2017} to explain the negative colour gradients in star-forming galaxies. {There is a feature in the surface SFR density of massive galaxies at $r\sim0.3r_{\rm e}$ with unusual lower SFR than the adjacent radial bins. This feature is also present in \citet{spindler2017} and \citet{belfiore2017}. However, an analysis at that level of detail goes beyond the scope of the present paper, and we do not attempt to reproduce it with our modelling.}
	
	To reproduce the difference between gas and stellar metallicity, the metal enrichment at early epochs needs to be suppressed. In the variable outflow scenario, the early transition {time} $t_{\rm out,e}$ of the outflow and the initial metal outflow fraction $f_{\rm out,i}$ are the two key parameters that regulate the suppression of early metal {enrichment} and thus the final stellar metallicity. 
	To reproduce the negative stellar metallicity gradient, the average metal outflow fraction has to be higher at larger radii. To achieve that, a higher $t_{\rm out}$ ($f_{\rm out,i}$) is required, which implies a longer (more extreme) initial phase of high metal outflow in the outer regions. 
	For example, for the most massive galaxy bin with the steepest stellar metallicity gradient, a steep gradient in $t_{\rm out}$ of 5Gyr$/r_{\rm e}$ is needed.
	
	This positive gradient in $t_{\rm out,e}$ is astrophysically plausible since the gravitational potential well in the outer regions of galaxies is relatively shallow. 
	{At the same time, star formation rates and hence the energy input are lower, though. Our results imply that the effect of the shallower potential well dominates, and the gas outflow at large radii must be more efficiently triggered. Gas outflow is also expected to last longer and to be more extreme at large radii during the evolution of a galaxy.}
	The steeper negative stellar metallicity gradient observed in more massive galaxies suggests that the evolution of massive galaxies is characterised by a steeper positive gradient in $t_{\rm out,e}$ and/or $f_{\rm out,i}$.

	\subsubsection{Comparison with NGC 628}
	In comparing the mass of metals in gaseous and stellar components in the galaxy NGC 628 \citet{belfiore2016a} found the average fraction of metals that has been lost to be $\sim50$ per cent. This is in qualitative agreement with our result when we adopt the empirical N2 calibration for the gas metallicity. However, \citet{sanchez2014a,belfiore2016a} found the loss fraction to be higher in the central region and to decrease toward larger radii, which is the opposite of what we conclude form the chemical evolution modelling presented here. This discrepancy is likely caused by the positive stellar metallicity gradient of this specific galaxy, which is not representative of the normal galaxy population in the local universe.
	
	\subsubsection{Drivers of the gas metallicity gradient}
	Different from the {\em stellar} metallicity gradient, the {\em gas} metallicity gradient is driven {by a combination of radially dependent inflow rate, i.e.\ the SFE, and recent metal outflow properties}. The inclusion of a second, constant period of (low) metal outflow fraction is {necessary for massive galaxies} {which show a flatter gas metallicity gradient than less massive galaxies}. If we do not consider the second constant period of (low) metal outflow fraction but instead assume a linear decline of the metal outflow fraction from the transition time $t_{\rm out,e}$ until today, then the {steep gradient of} $t_{\rm out,e}$ {in massive galaxies would affect} the negative gas metallicity gradient as well and the predicted gradient would be much steeper than the observation.
	
	
	\subsubsection{Effect of the gas metallicity calibration}
	To investigate how the discrepancy between different gas metallicity calibrations affects our results, we also tune the model to match the observed radial profiles in gas metallicity obtained through the empirical N2 method, alongside stellar metallicity, stellar mass surface density, and SFR surface density as before.
	
	Figure A1 in the Appendix shows the comparison between the observations and the predictions of the GCE model with variable metal outflow fraction. The parameters of the model shown in Figure A1 are listed in Table A1. 
	Since the empirical N2 method leads to a much lower gas metallicity measurement compared to the theoretical R23 method, the difference between gas and stellar metallicity becomes smaller. Nevertheless, as discussed in \textsection3.1, a variable outflow or variable IMF scenario is still needed to match the different slopes (i.e.\ dynamic range) of the gas and stellar metallicity radial profiles. Therefore, our main result is robust against the different gas metallicity calibrations, but the detail of the parameters derived differ quantitatively. With the smaller difference between gas and stars obtained through the N2 method, we can see from {Table A1} that a smaller early outflow transition time $t_{\rm out,e}$ (i.e.\ shorter duration time for the high outflow phase) and lower initial metal outflow fraction $f_{\rm out,i}$ is needed to match the data. Otherwise the conclusions remain the same.

	\subsection{Variable IMF model}
	
	\begin{figure*}
	    \centering
	    \includegraphics[width=13cm]{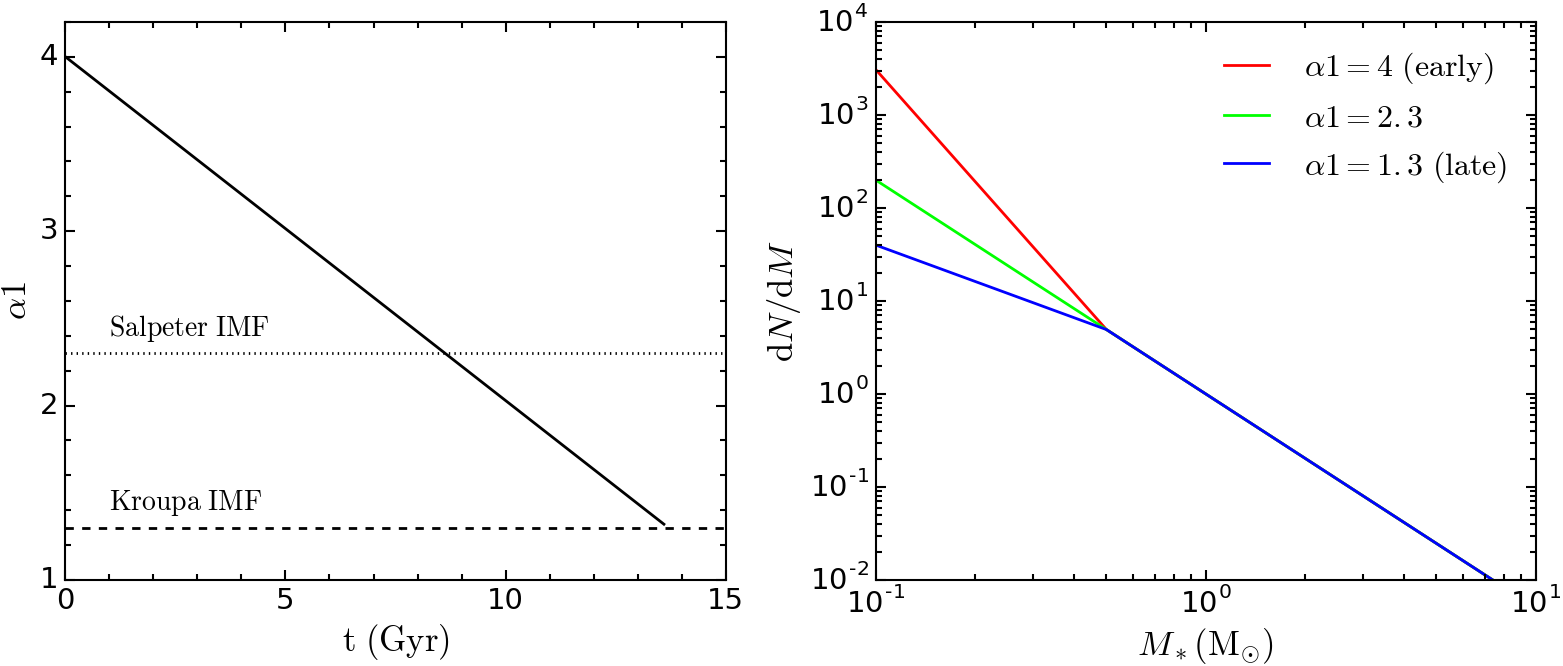}	
	    \caption{Schematic plot to illustrate the linear function adopted for the time evolution of the IMF slope at low mass end.}
	    \label{figure8}
	\end{figure*} 
	
	\begin{figure*}
		\centering
		\includegraphics[width=18cm]{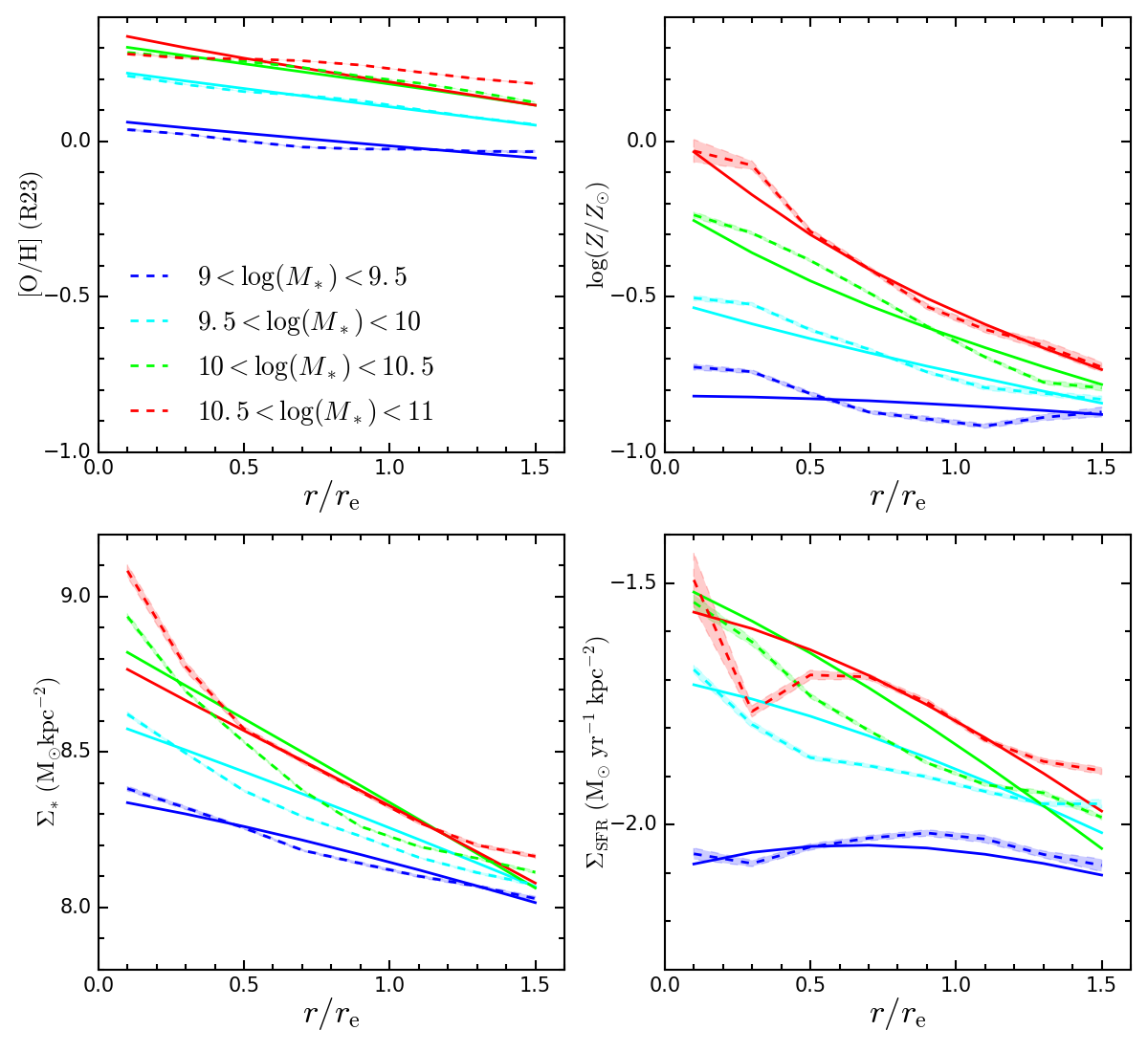}
		\caption{Similar figure to Figure 7 but showing the comparison between MaNGA data and a model with variable IMF slope (for detail see text). The parameters adopted in the variable IMF model are listed in Table 4.
		}
		\label{figure9}
	\end{figure*}
	
	\begin{table*}
		\caption{Parameter values of the time-dependent IMF model in Figure 9.}
		\label{Table4}
		\centering
		\begin{tabular}{l c c c c c c c c c c}
			\hline\hline
			mass bin & $A_{\rm ks}$ & $A_{\rm inf}$ & $\tau_{\rm inf}$ & $f_{\rm out}$ & $t_{\rm out,e}$ & $\alpha1$ &$\alpha1_{\rm i}$ & $\alpha2$ & $\alpha2_{\rm i}$ \\ 
			log(${\rm M_{\odot}}$) & - & ${\rm M}_{\odot}{\rm yr^{-1}}$ & Gyr & - & Gyr & - & - & - & - \\
			\hline
			$[9,9.5]$ & 0.5 & 1.4$\frac{r}{r_{\rm e}}e^{-0.8\frac{r}{r_{\rm e}}}$ & 3.5+1.7$\frac{r}{r_{\rm e}}$ & 0.26 & 13.7 & 1.3 & 4.2+0.5$\frac{r}{r_{\rm e}}$ & 2.3 & 2.3 \\
			$[9.5,10]$ & 0.6 & 3.5$\frac{r}{r_{\rm e}}e^{-1\frac{r}{r_{\rm e}}}$ & 4.5+1.5$\frac{r}{r_{\rm e}}$ & 0 & 13.7 & 1.3 & 4.0+1.1$\frac{r}{r_{\rm e}}$ & 2.3 & 2.3  \\
			$[10,10.5]$ & 1 & 12$\frac{r}{r_{\rm e}}e^{-1.4\frac{r}{r_{\rm e}}}$ & 4.5+1.5$\frac{r}{r_{\rm e}}$ & 0 & 13.7 & 1.3 & 3.2+1.6$\frac{r}{r_{\rm e}}$ & 2.3 & 2.3 \\
			$[10.5,11]$ & 1 & 23$\frac{r}{r_{\rm e}}e^{-1.3\frac{r}{r_{\rm e}}}$ & 4.5+1.5$\frac{r}{r_{\rm e}}$ & 0 & 13.7 & 1.3 & 2.6+2.0$\frac{r}{r_{\rm e}}$ & 2.3 & 2.3 \\
			\hline
		\end{tabular}\\  	
	\end{table*}
	
	As we found in Paper~I, another viable scenario to reproduce gas and stellar metallicities simultaneously requires a time-dependent IMF. Generally, a steep IMF at early times is needed to 
	suppress the metal enrichment and lower the stellar metallicity observed today. 
	In this scenario, we allow the initial IMF slope to vary but fix the present one to the double-power law Kroupa IMF. The IMF slope is assumed to change linearly with time. The slopes of the initial IMF at both the low-mass end ($\alpha1_{\rm i}$) and the high-mass end ($\alpha2_{\rm i}$) are free parameters. {Figure 8 illustrates the assumed time evolution of the IMF slope at the low mass end.}
	Based on this {double-power law} IMF, we explore two time-dependent IMF models with variations of the IMF slope at the low mass end and at the high mass end, separately.
	
	\subsubsection{Radially dependent IMF slope}
	Figure 9 shows the comparison between the MaNGA data and the model predictions for the model with variable IMF slope at the low mass end. 
	The comparison for the model with variable IMF slope at the high mass end is shown in Figure B1 in the Appendix.
	The line-style and colour scheme is the same as Figure 7. 
	Again, the model parameters are finely tuned to match the radial profiles of all four observables, gas and stellar metallicities as well as stellar mass surface density and SFR surface density. 
	Table 4 and Table B1 list the adopted parameters for the variable IMF models in Figure 9 and Figure B1, respectively.
	
	The initial IMF slope at the low mass end in the model is $\sim 5$ at large radii ($r\sim1.5r_{\rm e}$), nearly independently of the global mass of the galaxy. This is much steeper than a Kroupa IMF with a slope of 1.3 or even the Salpeter IMF with a slope of 2.3 \citep{salpeter1955}. The initial IMF slope required for the central region, instead, is highly dependent on the global mass of the galaxy, ranging from 4.2 at $10^9{\rm M_{\odot}}$ to 2.6 at $10^{10.5}{\rm M_{\odot}}$.
	
	
	As expected, also the GCE models with time dependent IMF slope are able to reproduce the radial profiles of gas and stellar metallicities simultaneously. {This is because less metals are produced with a steeper IMF. The latter implies a larger fraction of long-lived, low-mass stars that only return few metals to the ISM. This is consistent with the low outflow fractions since there is comparatively less supernova feedback from massive stars.} To match the steep negative stellar metallicity gradient, a positive gradient of average IMF slope is needed. We tune the average IMF slope by changing the initial IMF slope.
	
	\subsubsection{Dependence on galaxy mass}
	The steeper stellar metallicity gradient found in more massive galaxies requires a steeper gradient in the initial IMF slope. {This} is similar to our result in Paper~I, where we found that a steeper IMF was required in lower mass galaxies in order to match gas and stellar metallicities. However, this required a steeper initial IMF at larger radii and in less massive galaxies, which is the opposite of what is found in passive galaxies {based on the analysis of IMF-sensitive absorption-line indices} (\citealt{conroy2012,cappellari2012,van2017}; Parikh et al. in prep). 
	
	As can be seen in the top left-hand panel of Figure 9 the gas metallicity gradient in massive galaxies as predicted by the variable IMF model is mildly steeper than the observational data. This is due to the fact that
	changing the initial IMF slope not only changes the IMF slope at early times but also slightly changes the recent IMF slope because of the assumption of linear evolution. 
	
	\subsubsection{Effect of the gas metallicity calibration}
	We also investigate the impact of the gas metallicity calibration on the models presented in this section. The result is shown in Figure A2 in the Appendix. The finely tuned parameters are listed in Table A2. For simplicity, we only tune the model with variable IMF slope at the low mass end. Similar to the variable outflow scenario, when adopting the empirical N2 method, a time-dependent IMF is still needed to match the steep slope of the stellar metallicity gradient. As expected, due to the smaller difference between gas and stellar metallicity, a somewhat flatter initial IMF slope gradient is sufficient to match the observations.  
	
	Finally, it should be noted that, to match the relatively low gas metallicity derived through the empirical N2 method, we need either a strong {\sl present} metal outflow fraction of $\sim 60\%$, a steep {\sl present} IMF slope 
	or an extremely low KS law coefficient (10\% of the original value). These values are not supported by observations in the local universe where no clear signature of strong galactic metal outflows or steep IMF slopes
	is found in local star-forming galaxies \citep{concas2017,bastian2010}. However, given the uncertainties in metal yields through stellar mass loss, especially from Type-II supernova, such unrealistic parameter settings may not be required, if, for instance, the metal yields are overestimated.
	
	\section{Discussion}
	In this paper we analyse gradients of gas and stellar metallicity in star forming galaxies derived from MaNGA IFU observations. We show that gas metallicity is always higher than the average stellar metallicity, and that stellar metallicity gradients are signficantly steeper. As a consequence the discrepancy between gas and stellar metallicity is largest at large galaxy radii. As discussed in Paper~I, gas and stellar metallicity values can only be reconciled in chemical enrichment scenarios where metal enrichment is significantly suppressed at early times. We show that this can be achieved with either a time-dependent metal outflow with larger metal loading factors in galactic winds at early times, or through a time-dependent IMF with steeper IMF slopes at early times. The former manipulates {\em metal retention} implying that most metals are lost from the galaxy at early times, while the latter manipulates {\em metal production} implying that less metals are produced at early times.
	
	\subsection{Galactic wind vs IMF}
	Here we explore these scenarios and show that they can both explain the observed gas and stellar metallicity gradients. The retention model implies that early in the evolution metal outflow fractions must have been larger at large radii, as well as larger in lower mass galaxies. The production model, instead, suggests that the IMF must have been steeper at large radii and in lower mass galaxies in the past.

	{This second scenario could be caused by a metallicity dependence of the IMF implying that steeper IMFs would need to be generated in lower metallicity environments. This is not implausible, but clear evidence for such a behaviour is currently missing \citep{bastian2010,bate2014}. Possible IMF variations seen in simulations point to the opposite trends implying bottom-heavier IMFs in environments of higher turbulence and star-formation rate \citep{kroupa2013,chabrier2014}.} On top of this, a variable IMF as constrained here seems to enjoy little support from independent observational constraints on IMF variations in galaxies, though. While this subject is still controversially discussed in the literature \citep{smith2015}, IMF variations, if at all present, are seen in the opposite sense with the steepest IMF slopes generally inferred for the centres of more massive early-type galaxies \citep{conroy2012,cappellari2012,van2017,alton2017,li2017,parikh2018}. 
	
	{The variable IMF scenario explored here appears also in tension with the top-heavy IMF reported to be measured in highly star-forming galaxies by \citet{gunawardhana2011}. From an analysis of the {\ha\ equivalent width distribution} of galaxies as a function of optical colour, \citet{gunawardhana2011} find an SFR-dependent IMF with a flatter high-mass IMF in galaxies with more {vigorous} star formation. This trend implies a flatter IMF in more massive galaxies, which would lead to an overproduction of stellar metallicity in our model.} In the following we therefore focus on the wind model and the implications of time-dependent metal loading factors in galactic winds.

    	{Still, a combination of variable IMF and gas outflow may still be an adequate scenario. Chemical evolution models by \citet{recchi2015} show that both an IMF variation through implementation of the so-called integrated galactic initial mass function \citep[IGIMF,][]{kroupa2013} as well as additional in- and outflow of gas is required to reproduce the extremely low metallicities of low-mass galaxies.}
        
	\subsection{Metal loading in galactic winds}
	The mass and time dependence of the gas accretion rates assumed in our model are certainly plausible astrophysically with higher infall rates in more massive objects and a steeper subsequent decline. The gas accretion into a dark matter halo is not likely to be constant but declining with time as indicated by the cosmic evolution of SFR density \citep{madau2014}. However, like the gas accretion rate, the time scale of the decline may also depend on the galaxy mass and distance to the galaxy centre. Gas accretion probably declines faster in more massive galaxies or in more central regions of galaxies to shape the `flat' mass-SFR relation and the relative steeper gradient in mass surface density compared to the SFR surface density \citep[e.g.][]{renzini2015}. The mass dependence of the gas accretion time-scale leads to the downsizing scenario where more massive galaxies tend to have the majority of their stars formed at earlier times \citep{cowie1996,heavens2004,thomas2005}. The radial dependence fits to an inside-out growth scenario of star-forming galaxies and discs which results in the observed negative colour and stellar population gradients \citep{lian2017,goddard2017b}. 
	
	Our model suggests that galactic winds and their metal loading factors depend on cosmic time, galaxy mass and radius. But very importantly, it also implies relatively high mass and metal loading factors. These are indeed observed in starburst galaxies \citep{strickland2009}
    {and in good agreement with a study by \citet{Peeples2011} who show that efficient outflows are required to reproduce the observed correlation of gas metallicity and gas fraction with galaxy mass simultaneously. Also \citet{Peeples2011} find from their simulations that the metal expulsion efficiency must scale steeply with galaxy mass. Interestingly, \citet{taylor2015} show that metal outflow turns out to be small in a simulation of a massive galaxy, even if AGN driven. This result is consistent with the results of the present study, as strong metal outflows are indeed required in low-mass galaxies and at large radii}.
    
It is again reasonable to infer that deeper potential wells will lead to lower metal loading in galactic winds, which naturally leads to a dependence on both galaxy mass and galaxy radius in agreement with the model found here. Galaxies are more likely to lose their metals from their outskirts, and lower-mass galaxies are more likely to eject metals overall. In summary, the mass- and radially-dependent average metal outflow fraction shapes the MZR$_{\rm star}$ and the stellar metallicity gradient of local star-forming galaxies. Gas metallicity, instead, is generally determined by the recent evolutionary processes \citep{lilly2013}, including metal outflow, gas inflow, and star formation. A mass and radial dependence of one of these processes (or a combination of them) will then shape the MZR$_{\rm gas}$ and the gas metallicity gradient of star-forming galaxies.
	 	
	\subsection{Observational constraints from high redshift}
	A prediction of our model is that the metal outflow fraction, hence the metal loading factor of galactic winds, must have been higher at early times in the evolution of star forming galaxies. While local starburst galaxies do appear to exhibit supernova driven winds with relatively high metal loading factors, it would be interesting to check observationally whether star forming galaxies at higher redshifts are more efficient in ejecting metals than their low-$z$ counterparts. 
	
	{In terms of the metallicity gradient, our model predicts a steeper gas metallicity gradient at higher redshift. In other words, the gas metallicity gradient is expected to flatten with cosmic time. This is implied by the 
	relatively steeper gradient in stellar metallicity than the gas metallicity in local galaxies as presented here. Measurements of gas metallicity in planetary nebulae in the Milky Way agree with this picture and suggest that our Galaxy had a steeper gas metallicity gradient in the past up to $z\sim 1.5$ \citep{maciel2003}. 
	Observations at high redshifts, however, show more diverse results \citep[e.g.,][]{cresci2010,yuan2011,swinbank2012,jones2010,jones2013,wuyts2016}), with the slope of the gas metallicity gradient varying from negative to positive. A detailed and direct comparison between the prediction of our model and observations of the gas metallicity gradient at high redshift is beyond the scope of this paper and is {the} subject of future work. 
	}

	\subsection{Cosmological models}
	While the model presented here suggests a scenario to explain gas and stellar metallicities in galaxies, it does not explore the physical origin of it. As discussed above the scenarios are astrophysically plausible, but it will be important to investigate whether the high metal outflow fractions assumed here can be accommodated in cosmological simulations of galaxy formation \citep[e.g.,][]{calura2009,yates2012,taylor2015,guo2016,ma2017,rossi2017,taylor2017}.

	\section{Conclusions}
	In this work, we investigate the radial distribution of gas metallicity, stellar metallicity, mass surface density, and SFR surface density of local star-forming galaxies using IFU data from the SDSS-IV/MaNGA survey.   
	The metallicity in the gaseous component of these star-forming galaxies is generally higher than the metallicity in the stellar component, especially in low mass galaxies and at large radii. 
	When looking at the radial distribution, we find that most of the star-forming galaxies in our sample present a negative gradient in both gas and stellar metallicity. However, the gradient in stellar metallicity is much steeper than the gas metallicity. By dividing the galaxy sample into four mass bins, we also investigate the mass-dependence of these gradients. 
	
	It turns out that the radial profile of the {\em stellar} metallicity steepens considerably in massive galaxies while no clear dependence on the global mass is found for the {\em gas} metallicity radial profile. The steepening of the stellar metallicity gradient with mass is mostly due to the enhanced stellar metallicity in the central region of massive galaxies, suggesting that the chemical evolution of the galaxy central region is dependent on the global mass. On the contrary, the stellar metallicity at large radii of galaxies tend to converge which implies an universal chemical evolution history of the outer disc components of galaxies.      
	In terms of mass and SFR surface density, the 
	radial distribution of the mass surface density is power-like and is steeper than that of the 
	SFR surface density, suggesting that the gas inflow rate declines faster in the central regions of galaxies which is consistent with the inside-out growth scenario of disc galaxies. 
	
	To investigate the origin of the negative gradient in gas and stellar metallicity in a galaxy formation and evolution framework, we use a chemical evolution model to reproduce the observed radial distribution of galaxy properties, including the gas metallicity, stellar metallicity, mass surface density, and SFR surface density.
	We assume an initial gas inflow surface density that declines exponentially with radius to match the power law-like radial profile of the mass surface density.
	Following our results from Paper~I, we show that either a time-dependent metal outflow fraction (proportional to the mass loading factor) with higher fraction at early times or a time-dependent IMF slope with steeper IMF at early times is needed to reproduce the different metallicity in the gaseous and stellar components and the different slopes in their radial profiles. We also explore the effect of using different gas metallicity calibrations. While the lower gas metallicities obtained by some empirical methods reduce the absolute difference in gas and stellar metallicity, these two scenarios are still needed to match the different slopes in the radial profiles of gas and stellar metallicity. 
	
	Based on these two scenarios and using the chemical evolution model developed in Paper~I, for the first time we reproduce the radial distribution of gas and stellar metallicity simultaneously. 
	In the time-dependent outflow model, the disc components of galaxies have experienced a higher fraction of metal loss than the central region on average. The negative stellar metallicity gradient is driven by the positive gradient in the average fraction of metal loss which could be achieved by a positive gradient in the early metal outflow transition time $t_{\rm out,e}$ and/or initial metal outflow fraction $f_{\rm out,i}$. 
	The steeper stellar gradient observed in more massive galaxies suggests a steeper gradient in the average fraction of metal loss.

	In the time-dependent IMF scenario the stellar metallicity gradient is regulated by a positive gradient in the average IMF slope which is achieved by a positive gradient in the initial IMF slope with steeper initial IMF slopes at larger radii.
	In both scenarios, the gas metallicity is regulated by the recent evolutionary processes while the stellar metallicity is additionally regulated by the processes at early epochs. 
	In conclusion from this study and Paper I, we find that either the metal loading factor in galactic winds must have been high in the past with the longest duration time and the highest loading factor of this outflow phase at the largest radii and lowest galaxy masses, or the IMF slope must have been steeper in the past, again with the strongest variation at the largest radii and the lowest mass galaxies.
	
	A direct consequence of our galactic wind model is that the metal outflow fraction, hence the metal loading factor of galactic winds, must have been higher at higher redshift, a prediction that would be interesting to check observationally. Finally, {while} the model presented here suggests a scenario to explain gas and stellar metallicities in galaxies, the physical origin remains yet to be explored in the framework of cosmological simulations of galaxy formation.

	\section*{Acknowledgements}
	
    The Science, Technology and Facilities Council is acknowledged for support through the Consolidated Grant ‘Cosmology and Astrophysics at Portsmouth’, ST/N000668/1. Numerical computations were done on the Sciama High Performance Compute (HPC) cluster which is supported by the ICG, SEPnet and the University of Portsmouth.
    
	Funding for the Sloan Digital Sky Survey IV has been provided by the Alfred P. Sloan Foundation, the U.S. Department of Energy Office of Science, and the Participating Institutions. SDSS acknowledges support and resources from the Center for High-Performance Computing at the University of Utah. The SDSS web site is www.sdss.org.
	
	SDSS is managed by the Astrophysical Research Consortium for the Participating Institutions of the SDSS Collaboration including the Brazilian Participation Group, the Carnegie Institution for Science, Carnegie Mellon University, the Chilean Participation Group, the French Participation Group, Harvard-Smithsonian Center for Astrophysics, Instituto de Astrofísica de Canarias, The Johns Hopkins University, Kavli Institute for the Physics and Mathematics of the Universe (IPMU) / University of Tokyo, Lawrence Berkeley National Laboratory, Leibniz Institut für Astrophysik Potsdam (AIP), Max-Planck-Institut für Astronomie (MPIA Heidelberg), Max-Planck-Institut für Astrophysik (MPA Garching), Max-Planck-Institut für Extraterrestrische Physik (MPE), National Astronomical Observatories of China, New Mexico State University, New York University, University of Notre Dame, Observatório Nacional / MCTI, The Ohio State University, Pennsylvania State University, Shanghai Astronomical Observatory, United Kingdom Participation Group, Universidad Nacional Autónoma de México, University of Arizona, University of Colorado Boulder, University of Oxford, University of Portsmouth, University of Utah, University of Virginia, University of Washington, University of Wisconsin, Vanderbilt University, and Yale University.
	

\begin{thebibliography}{60}  
		\bibitem[Abolfathi et al.(2017)]{abolfathi2017} Abolfathi, B., Aguirre, V.~S., Almeida, A., et al.\ 2017, submitted
		
		\bibitem[Alton et al.(2017)]{alton2017} Alton, P.~D., Smith, R.~J., \& Lucey, J.~R.\ 2017, MNRAS, 468, 1594 
		
		\bibitem[\protect\citeauthoryear{Andrews 
			\& Martini}{2013}]{andrews2013} Andrews, B.~H., \& Martini, P.\ 2013, ApJ, 765, 140
		
		\bibitem[\protect\citeauthoryear{Asplund et 
			al.}{2009}]{asplund2009} Asplund, M., Grevesse, N., Sauval, A.~J., \& Scott, P.\ 2009, ARA\&A, 47, 481
		
		\bibitem[Bacon et al.(2001)]{bacon2001} Bacon, R., Copin, Y., Monnet, G., et al.\ 2001, MNRAS, 326, 23 
		
		\bibitem[\protect\citeauthoryear{Baldwin, Phillips \& Terlevich}{Baldwin et al.}{1981}]{baldwin1981} Baldwin, J.~A., 
		Phillips, M.~M., \& Terlevich, R.\ 1981, PASP, 93, 5	
        
        \bibitem[Bate(2014)]{bate2014} Bate, M.~R.\ 2014, MNRAS, 442, 285 
		
		\bibitem[Bastian et al.(2010)]{bastian2010} Bastian, N., Covey, K.~R., \& Meyer, M.~R.\ 2010, ARA\&A, 48, 339 
		
		\bibitem[Belfiore et al.(2016)]{belfiore2016a} Belfiore, F., Maiolino, R., \& Bothwell, M.\ 2016, MNRAS, 455, 1218 
		
		
		\bibitem[Belfiore et al.(2017)]{belfiore2017} Belfiore, F., Maiolino, R., Tremonti, C., et al.\ 2017, arXiv:1703.03813 
		
		\bibitem[Blanton et al.(2017)]{blanton2017} Blanton, M.~R., Bershady, M.~A., Abolfathi, B., et al.\ 2017, AJ, 154, 28 
		
		\bibitem[Blanton et al.(2011)]{blanton2011} Blanton, M.~R., Kazin, E., Muna, D., Weaver, B.~A., \& Price-Whelan, A.\ 2011, AJ, 142, 31 
		
		\bibitem[Bundy et al.(2015)]{bundy2015} Bundy, K., Bershady, M.~A., Law, D.~R., et al.\ 2015, ApJ, 798, 7 
		
		\bibitem[Calura et al.(2009)]{calura2009} Calura, F., Pipino, A., Chiappini, C., Matteucci, F., \& Maiolino, R.\ 2009, A\&A, 504, 373 
		
		\bibitem[Cappellari \& Copin(2003)]{cappellari2003} Cappellari, M., \& Copin, Y.\ 2003, MNRAS, 342, 345
		
		\bibitem[Cappellari et al.(2011)]{cappellari2011} Cappellari, M., Emsellem, E., Krajnovi{\'c}, D., et al.\ 2011, MNRAS, 413, 813  
		
		\bibitem[Cappellari et al.(2012)]{cappellari2012} Cappellari, M., McDermid, R.~M., Alatalo, K., et al.\ 2012, Nature, 484, 485 
		
		\bibitem[Cardelli et al.(1989)]{ccm89} Cardelli, J.~A., Clayton, G.~C., \& Mathis, J.~S.\ 1989, ApJ, 345, 245 
		
		\bibitem[Carollo et al.(1993)]{carollo1993} Carollo, C.~M., Danziger, I.~J., \& Buson, L.\ 1993, MNRAS, 265, 553 
        
        \bibitem[Chabrier et al.(2014)]{chabrier2014} Chabrier, G., Hennebelle, P., \& Charlot, S.\ 2014, APJ, 796, 75 
		
		\bibitem[Chiappini et al.(2001)]{chiappini2001} Chiappini, C., Matteucci, F., \& Romano, D.\ 2001, ApJ, 554, 1044 
		
		\bibitem[Concas et al.(2017)]{concas2017} Concas, A., Popesso, P., Brusa, M., et al.\ 2017, arXiv:1701.06569
		
		\bibitem[Conroy \& van Dokkum(2012)]{conroy2012} Conroy, C., \& van Dokkum, P.~G.\ 2012, ApJ, 760, 71 
		
		\bibitem[Cowie et al.(1996)]{cowie1996} Cowie, L.~L., Songaila, A., Hu, E.~M., \& Cohen, J.~G.\ 1996, AJ, 112, 839 
		
		\bibitem[Cresci et al.(2010)]{cresci2010} Cresci, G., Mannucci, F., Maiolino, R., et al.\ 2010, Nature, 467, 811 
		
		
		\bibitem[Croom et al.(2012)]{croom2012} Croom, S.~M., Lawrence, J.~S., Bland-Hawthorn, J., et al.\ 2012, MNRAS, 421, 872 
		
		\bibitem[Davies et al.(1993)]{davies1993} Davies, R.~L., Sadler, E.~M., \& Peletier, R.~F.\ 1993, MNRAS, 262, 650  
		
		\bibitem[De Rossi et al.(2017)]{rossi2017} De Rossi, M.~E., Bower, R.~G., Font, A.~S., Schaye, J., \& Theuns, T.\ 2017, arXiv:1704.00006 
		
		\bibitem[Drory et al.(2015)]{drory2015} Drory, N., MacDonald, N., Bershady, M.~A., et al.\ 2015, AJ, 149, 77 
		
		\bibitem[Ellison et al.(2008)]{ellison2008} Ellison, S.~L., Patton, D.~R., Simard, L., \& McConnachie, A.~W.\ 2008, ApJL, 672, L107 
		
		
		\bibitem[Fu et al.(2009)]{fu2009} Fu, J., Hou, J.~L., Yin, J., \& Chang, R.~X.\ 2009, ApJ, 696, 668 
		
		
		\bibitem[Gallazzi et al.(2005)]{gallazzi2005} Gallazzi, A., Charlot, S., Brinchmann, J., White, S.~D.~M., \& Tremonti, C.~A.\ 2005, MNRAS, 362, 41 
		
		\bibitem[Gibson et al.(2013)]{gibson2013} Gibson, B.~K., Pilkington, K., Brook, C.~B., Stinson, G.~S., \& Bailin, J.\ 2013, A\&A, 554, A47 
		
		\bibitem[Goddard et al.(2017a)]{goddard2017a} Goddard, D., Thomas, D., Maraston, C., et al.\ 2017a, MNRAS, 465, 688 
		
		
		\bibitem[Goddard et al.(2017b)]{goddard2017b} Goddard, D., Thomas, D., Maraston, C., et al.\ 2017b, MNRAS, 466, 4731 
		
		
		\bibitem[Gonz{\'a}lez Delgado et al.(2015)]{gonzalez2015} Gonz{\'a}lez Delgado, R.~M., Garc{\'{\i}}a-Benito, R., P{\'e}rez, E., et al.\ 2015, A\&A, 581, A103 
		
		\bibitem[Gunawardhana et al.(2011)]{gunawardhana2011} Gunawardhana, M.~L.~P., Hopkins, A.~M., Sharp, R.~G., et al.\ 2011, MNRAS, 415, 1647 
		
		\bibitem[Guo et al.(2016)]{guo2016} Guo, Q., Gonzalez-Perez, V., Guo, Q., et al.\ 2016, MNRAS, 461, 3457 
		
		\bibitem[Heavens et al.(2004)]{heavens2004} Heavens, A., Panter, B., Jimenez, R., \& Dunlop, J.\ 2004, Nature, 428, 625 
		
		\bibitem[Ho et al.(2015)]{ho2015} Ho, I.-T., Kudritzki, R.-P., Kewley, L.~J., et al.\ 2015, MNRAS, 448, 2030 
		
		\bibitem[Iwamoto et al.(1999)]{iwamoto1999} Iwamoto, K., Brachwitz, F., Nomoto, K., et al.\ 1999, ApJS, 125, 439 
		
		\bibitem[Jones et al.(2010)]{jones2010} Jones, T., Ellis, R., Jullo, E., \& Richard, J.\ 2010, ApJL, 725, L176 
		
		\bibitem[Jones et al.(2013)]{jones2013} Jones, T., Ellis, R.~S., Richard, J., \& Jullo, E.\ 2013, ApJ, 765, 48 
		
		\bibitem[J{\o}rgensen(1999)]{jorgensen1999} J{\o}rgensen, I.\ 1999, MNRAS, 306, 607 
		
		\bibitem[\protect\citeauthoryear{Kauffmann et al.}{2003}]{kauffmann2003} Kauffmann, G. 
		et al.\ 2003, MNRAS, 346, 1055 
		
		\bibitem[\protect\citeauthoryear{Kennicutt}{1998}]{kennicutt1998} Kennicutt, R.~C., Jr.\ 1998, ARA\&A, 36, 189 
		
		\bibitem[\protect\citeauthoryear{Kewley et al.}{2001}]{kewley2001} Kewley, L.~J., Dopita, 
		M.~A., Sutherland, R.~S., Heisler, C.~A., 
		\& Trevena, J.\ 2001, ApJ, 556, 121	
		
		\bibitem[Kewley \& Dopita(2002)]{kewley2002} Kewley, L.~J., \& Dopita, M.~A.\ 2002, ApJS, 142, 35 
		
		\bibitem[\protect\citeauthoryear{Kewley 
			\& Ellison}{2008}]{kewley2008} Kewley, L.~J., \& Ellison, S.~L.\ 2008, ApJ, 681, 1183
		
		\bibitem[\protect\citeauthoryear{Kroupa}{2001}]{kroupa2001} Kroupa, P.\ 2001, MNRAS, 322, 231
		
		\bibitem[Kroupa et al.(2013)]{kroupa2013} Kroupa, P., Weidner, C., Pflamm-Altenburg, J., et al.\ 2013, Planets, Stars and Stellar Systems.~Volume 5: Galactic Structure and Stellar Populations, 5, 115
		
		\bibitem[Kuntschner et al.(2010)]{kuntschner2010} Kuntschner, H., Emsellem, E., Bacon, R., et al.\ 2010, MNRAS, 408, 97 
		
		\bibitem[Lara-L{\'o}pez et al.(2010)]{lopez2010} Lara-L{\'o}pez, M.~A., Cepa, J., Bongiovanni, A., et al.\ 2010, A\&A, 521, L53 
		
		\bibitem[Law et al.(2015)]{law2015} Law, D.~R., Yan, R., Bershady, M.~A., et al.\ 2015, AJ, 150, 19 
		
		\bibitem[Law et al.(2016)]{law2016} Law, D., et al.\ 2016, ApJ, in press
		
		\bibitem[\protect\citeauthoryear{Lequeux et 
			al.}{1979}]{lequeux1979} Lequeux, J., Peimbert, M., Rayo, J.~F., Serrano, A., \& Torres-Peimbert, S.\ 1979, A\&A, 80, 155
		
		\bibitem[Li et al.(2017)]{li2017} Li, H., Ge, J., Mao, S., et al.\ 2017, ApJ, 838, 77 

		\bibitem[Lian et al.(2016)]{lian2016} Lian, J., Yan, R., Zhang, K., \& Kong, X.\ 2016, ApJ, 832, 29 
		
		\bibitem[Lian et al.(2017)]{lian2017} Lian, J., Thomas, D., Maraston, C., et al.\ 2017, arXiv:1710.11135 
		
		\bibitem[Lilly et al.(2013)]{lilly2013} Lilly, S.~J., Carollo, C.~M., Pipino, A., Renzini, A., \& Peng, Y.\ 2013, ApJ, 772, 119 	
		
		\bibitem[Ma et al.(2017)]{ma2017} Ma, X., Hopkins, P.~F., Feldmann, R., et al.\ 2017, MNRAS, 466, 4780 
		
		\bibitem[Maciel et al.(2003)]{maciel2003} Maciel, W.~J., Costa, R.~D.~D., \& Uchida, M.~M.~M.\ 2003, A\&A, 397, 667 
		
		\bibitem[\protect\citeauthoryear{Maiolino et al.}{2008}]{maiolino2008} Maiolino, R. et al.\ 2008, A\&A, 488, 463
		
		\bibitem[\protect\citeauthoryear{Mannucci et al.}{2010}]{mannucci2010} Mannucci, F., Cresci, 
		G., Maiolino, R., Marconi, A., \& Gnerucci, A.\ 2010, MNRAS, 408, 2115
		
		\bibitem[Marino et al.(2013)]{marino2013} Marino, R.~A., Rosales-Ortega, F.~F., S{\'a}nchez, S.~F., et al.\ 2013, A\&A, 559, A114 
		
		\bibitem[Martin et al.(2005)]{martin2005} Martin, D.~C., Fanson, J., Schiminovich, D., et al.\ 2005, ApJL, 619, L1 
		
		\bibitem[Maraston \& Str{\"o}mb{\"a}ck(2011)]{maraston2011} Maraston, C., \& Str{\"o}mb{\"a}ck, G.\ 2011, MNRAS, 418, 2785
		
		\bibitem[Madau \& Dickinson(2014)]{madau2014} Madau, P., \& Dickinson, M.\ 2014, ARA\&A, 52, 415 
		
		\bibitem[McGaugh(1991)]{m91} McGaugh, S.~S.\ 1991, ApJ, 380, 140 
		
		\bibitem[Mehlert et al.(2003)]{mehlert2003} Mehlert, D., Thomas, D., Saglia, R.~P., Bender, R., \& Wegner, G.\ 2003, A\&A, 407, 423 
		
		
		\bibitem[Moll{\'a} et al.(1997)]{molla1997} Moll{\'a}, M., Ferrini, F., \& D{\'{\i}}az, A.~I.\ 1997, ApJ, 475, 519 
		
		
		\bibitem[Mott et al.(2013)]{mott2013} Mott, A., Spitoni, E., \& Matteucci, F.\ 2013, MNRAS, 435, 2918 
		
		\bibitem[Moustakas et al.(2010)]{moustakas2010} Moustakas, J., Kennicutt, R.~C., Jr., Tremonti, C.~A., et al.\ 2010, ApJS, 190, 233-266 
		
		\bibitem[Panter et al.(2008)]{panter2008} Panter, B., Jimenez, R., Heavens, A.~F., \& Charlot, S.\ 2008, MNRAS, 391, 1117 
        
        		\bibitem[Parikh et al.(2018)]{parikh2018} Parikh, T., Thomas, D., Maraston, C., et al.\ 2018, MNRAS, submitted
		
		\bibitem[Peeples \& Shankar(2011)]{Peeples2011} Peeples, M.~S., \& Shankar, F.\ 2011, MNRAS, 417, 2962 
		
		\bibitem[P{\'e}rez-Montero et al.(2016)]{perez2016} P{\'e}rez-Montero, E., Garc{\'{\i}}a-Benito, R., V{\'{\i}}lchez, J.~M., et al.\ 2016, A\&A, 595, A62 
		
		\bibitem[\protect\citeauthoryear{Pettini 
			\& Pagel}{2004}]{pettini2004} Pettini, M., \& Pagel, B.~E.~J.\ 2004, MNRAS, 348, L59 
		
		\bibitem[\protect\citeauthoryear{Pilyugin 
			\& Thuan}{2005}]{pilyugin2005} Pilyugin, L.~S., \& Thuan, T.~X.\ 2005, ApJ, 631, 231
		
		\bibitem[Pilkington et al.(2012)]{pilkington2012} Pilkington, K., Few, C.~G., Gibson, B.~K., et al.\ 2012, A\&A, 540, A56 
		
		\bibitem[Pipino et al.(2010)]{pipino2010} Pipino, A., D'Ercole, A., Chiappini, C., \& Matteucci, F.\ 2010, MNRAS, 407, 1347 
		
		
		\bibitem[Portinari et al.(1998)]{portinari1998} Portinari, L., Chiosi, C., \& Bressan, A.\ 1998, A\&A, 334, 505 
		
		\bibitem[Prantzos \& Boissier(2000)]{prantzos2000} Prantzos, N., \& Boissier, S.\ 2000, MNRAS, 313, 338 
		
		\bibitem[Rahimi et al.(2011)]{rahimi2011} Rahimi, A., Kawata, D., Allende Prieto, C., et al.\ 2011, MNRAS, 415, 1469
		
		\bibitem[Recchi \& Kroupa(2015)]{recchi2015} Recchi, S., \& Kroupa, P.\ 2015, MNRAS, 446, 4168  
		
		\bibitem[Renzini \& Peng(2015)]{renzini2015} Renzini, A., \& Peng, Y.-j.\ 2015, ApJL, 801, L29 
		
		\bibitem[S{\'a}nchez-Bl{\'a}zquez et al.(2007)]{sanchez2007} S{\'a}nchez-Bl{\'a}zquez, P., Forbes, D.~A., Strader, J., Brodie, J., \& Proctor, R.\ 2007, MNRAS, 377, 759 
		
		\bibitem[S{\'a}nchez-Menguiano et al.(2016)]{sanchez2016} S{\'a}nchez-Menguiano, L., S{\'a}nchez, S.~F., P{\'e}rez, I., et al.\ 2016, A\&A, 587, A70 
		
		\bibitem[S{\'a}nchez et al.(2012)]{sanchez2012} S{\'a}nchez, S.~F., Kennicutt, R.~C., Gil de Paz, A., et al.\ 2012, A\&A, 538, A8 
		
		\bibitem[S{\'a}nchez-Bl{\'a}zquez et al.(2014)]{sanchez2014a} S{\'a}nchez-Bl{\'a}zquez, P., Rosales-Ortega, F., Diaz, A., \& S{\'a}nchez, S.~F.\ 2014, MNRAS, 437, 1534 
		
		\bibitem[S{\'a}nchez et al.(2014)]{sanchez2014} S{\'a}nchez, S.~F., Rosales-Ortega, F.~F., Iglesias-P{\'a}ramo, J., et al.\ 2014, A\&A, 563, A49 
		
		\bibitem[Sch{\"o}nrich \& McMillan(2017)]{schonrich2017} Sch{\"o}nrich, R., \& McMillan, P.~J.\ 2017, MNRAS, 467, 1154 
		
		\bibitem[Salpeter(1955)]{salpeter1955} Salpeter, E.~E.\ 1955, ApJ, 121, 161 
		
		\bibitem[\protect\citeauthoryear{Shi et al.}{2005}]{shi2005} Shi, F., Kong, X., Li, C., \& Cheng, F.~Z.\ 2005, A\&A, 437, 849
		
		\bibitem[Smith et al.(2015)]{smith2015} Smith, R.~J., Lucey, J.~R., \& Conroy, C.\ 2015, MNRAS, 449, 3441 
		
		\bibitem[Spindler et al.(2017)]{spindler2017} Spindler, A., Wake, D., Belfiore, F., et al.\ 2017, arXiv:1710.05049 
		
		\bibitem[Spolaor et al.(2010)]{spolaor2010} Spolaor, M., Kobayashi, C., Forbes, D.~A., Couch, W.~J., \& Hau, G.~K.~T.\ 2010, MNRAS, 408, 272
		
		\bibitem[Strickland \& Heckman(2009)]{strickland2009} Strickland, D.~K., \& Heckman, T.~M.\ 2009, ApJ, 697, 2030 
		
		\bibitem[Swinbank et al.(2012)]{swinbank2012} Swinbank, A.~M., Sobral, D., Smail, I., et al.\ 2012, MNRAS, 426, 935 
		
		\bibitem[Taylor \& Kobayashi(2015)]{taylor2015} Taylor, P., \& Kobayashi, C.\ 2015, MNRAS, 452, L59 
		
		
		\bibitem[Taylor \& Kobayashi(2017)]{taylor2017} Taylor, P., \& Kobayashi, C.\ 2017, MNRAS, 471, 3856 
		
		\bibitem[Thomas et al.(1999)]{thomas1999} Thomas, D., Greggio, L., \& Bender, R.\ 1999, MNRAS, 302, 537 
		
		\bibitem[Thomas et al.(2005)]{thomas2005} Thomas, D., Maraston, C., Bender, R., \& Mendes de Oliveira, C.\ 2005, ApJ, 621, 673 
		
		\bibitem[Thomas et al.(2010)]{thomas2010} Thomas, D., Maraston, C., Schawinski, K., Sarzi, M., \& Silk, J.\ 2010, MNRAS, 404, 1775 
		
		\bibitem[\protect\citeauthoryear{Tremonti et al.}{2004}]{tremonti2004} Tremonti, C.~A. 
		et al.\ 2004, ApJ, 613, 898
		
		\bibitem[Vila-Costas \& Edmunds(1992)]{vila-costas1992} Vila-Costas, M.~B., \& Edmunds, M.~G.\ 1992, MNRAS, 259, 121 
		
		\bibitem[Zaritsky et al.(1994)]{zaritsky1994} Zaritsky, D., Kennicutt, R.~C., Jr., \& Huchra, J.~P.\ 1994, ApJ, 420, 87 
		
		\bibitem[van Dokkum et al.(2017)]{van2017} van Dokkum, P., Conroy, C., Villaume, A., Brodie, J., \& Romanowsky, A.~J.\ 2017, ApJ, 841, 68 
		
		\bibitem[van Zee et al.(1998)]{vanzee1998} van Zee, L., Salzer, J.~J., Haynes, M.~P., O'Donoghue, A.~A., \& Balonek, T.~J.\ 1998, AJ, 116, 2805 
		
		\bibitem[Ventura et al.(2013)]{ventura2013} Ventura, P., Di Criscienzo, M., Carini, R., \& D'Antona, F.\ 2013, MNRAS, 431, 3642 
		
		\bibitem[Wake et al.(2017)]{wake2017} Wake, D.~A., Bundy, K., Diamond-Stanic, A.~M., et al.\ 2017, AJ, 154, 86
		
		\bibitem[Westfall et al, in prep]{westfall2017} Westfall, K., et al., in prep
		
		\bibitem[Wilkinson et al.(2015)]{wilkinson2015} Wilkinson, D.~M., Maraston, C., Thomas, D., et al.\ 2015, MNRAS, 449, 328 
		
		\bibitem[Wilkinson et al.(2017)]{wilkinson2017} Wilkinson, D.~M., Maraston, C., Goddard, D., Thomas, D., \& Parikh, T.\ 2017, MNRAS, 472, 4297 
		
		\bibitem[Woosley \& Weaver(1995)]{woosley1995} Woosley, S.~E., \& Weaver, T.~A.\ 1995, ApJS, 101, 181 
		
		\bibitem[Wuyts et al.(2016)]{wuyts2016} Wuyts, E., Wisnioski, E., Fossati, M., et al.\ 2016, ApJ, 827, 74 
		
		\bibitem[Yan et al.(2016a)]{yan2016a} Yan, R., Tremonti, C., Bershady, M.~A., et al.\ 2016, AJ, 151, 8 
		
		\bibitem[Yan et al.(2016b)]{yan2016b} Yan, R., Bundy, K., Law, D.~R., et al.\ 2016, AJ, 152, 197 
		
		\bibitem[Yates et al.(2012)]{yates2012} Yates, R.~M., Kauffmann, G., \& Guo, Q.\ 2012, MNRAS, 422, 215 
		
		\bibitem[Yuan et al.(2011)]{yuan2011} Yuan, T.-T., Kewley, L.~J., Swinbank, A.~M., Richard, J., \& Livermore, R.~C.\ 2011, ApJL, 732, L14 
		
		\bibitem[Zheng et al.(2017)]{zheng2017} Zheng, Z., Wang, H., Ge, J., et al.\ 2017, MNRAS, 465, 4572
		
	\end{thebibliography}

	\appendix
	\section{Gas metallicity by N2 method}
	Here we present the finely tuned viable chemical evolution models with time-dependent outflow fraction and IMF slope to match the observed the radial profile of gas metallicity (N2 method), stellar metallicity, mass surface density, and SFR surface density 
	as shown in Figure A1 and Figure A2, respectively. 
	\begin{figure*}
		\centering
		\includegraphics[width=18cm]{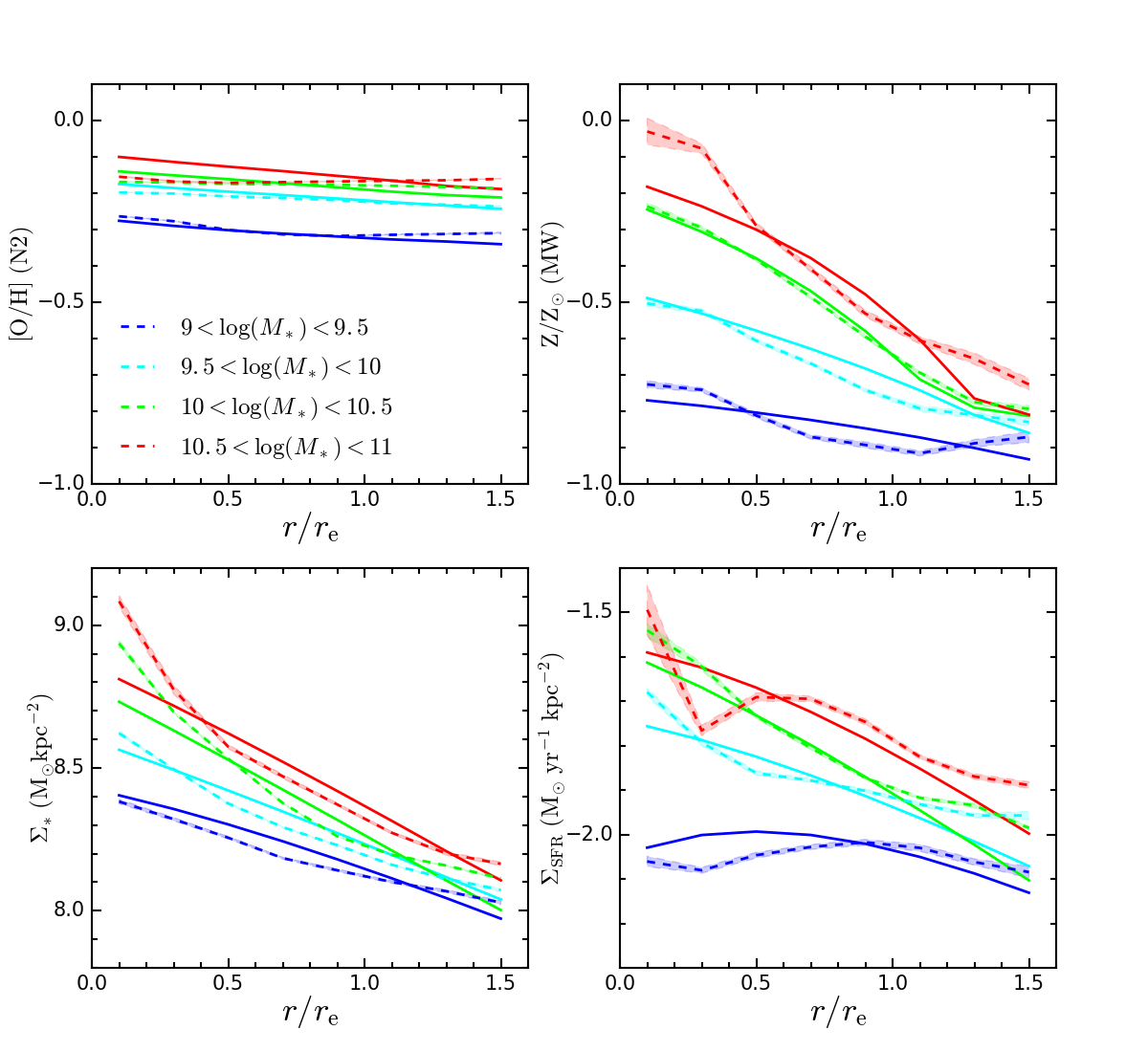}
		\caption{Similar to Figure 7 except that the gas metallicity is derived by empirical N2 method rather than the theoretical R23 method.     
		}
		\label{figureA1}
	\end{figure*}
	
	\begin{table*}
		\caption{Parameter value of the time-dependent outflow model in Figure A1.}
		\label{TableA1}
		\centering
		\begin{tabular}{l c c c c c c c c c c c c}
			\hline\hline
			mass bin & $A_{\rm ks}$ & $A_{\rm inf}$ & $\tau_{\rm inf}$ & $f_{\rm out,i}$ & $f_{\rm out,p}$ & $t_{\rm out,e}$ &$t_{\rm out,r}$ & $\alpha1$ &$\alpha1_{\rm i}$ & $\alpha2$&$\alpha2_{\rm i}$ \\ 
			log(${\rm M_{\odot}}$) & - & ${\rm M}_{\odot}{\rm yr^{-1}}$ & Gyr & - & - & Gyr & Gyr & - & - & - & - \\
			\hline
			$[9,9.5]$ & 1 & 1.3$\frac{r}{r_{\rm e}}e^{-1\frac{r}{r_{\rm e}}}$ & 4.4+2.8$\frac{r}{r_{\rm e}}$ & 0.93+0.02$\frac{r}{r_{\rm e}}$ & 0.73 & 2+2$\frac{r}{r_{\rm e}}$ & 13.7 & 1.3 & 1.3 & 2.3 & 2.3 \\
			$[9.5,10]$ & 1 & 2.8$\frac{r}{r_{\rm e}}e^{-1\frac{r}{r_{\rm e}}}$  & 5.5+1.8$\frac{r}{r_{\rm e}}$ & 0.83+0.12$\frac{r}{r_{\rm e}}$ & 0.66 & 2+$\frac{r}{r_{\rm e}}$ & 13.2 & 1.3 & 1.3 & 2.3 & 2.3 \\
			$[10,10.5]$ & 1 & 8$\frac{r}{r_{\rm e}}e^{-1.3\frac{r}{r_{\rm e}}}$ & 5.5+1.8$\frac{r}{r_{\rm e}}$ & 0.61+33$\frac{r}{r_{\rm e}}$ & 0.66 & 1+2$\frac{r}{r_{\rm e}}$ & 11.7 & 1.3 & 1.3 & 2.3 & 2.3 \\
			$[10.5,11]$ & 1 & 24$\frac{r}{r_{\rm e}}e^{-1.3\frac{r}{r_{\rm e}}}$ & 5+2.2$\frac{r}{r_{\rm e}}$ & 0.54+0.35$\frac{r}{r_{\rm e}}$ & 0.64 & 3$\frac{r}{r_{\rm e}}$ & 11.7 & 1.3 & 1.3 & 2.3 & 2.3 \\
			\hline
		\end{tabular}\\  	
	\end{table*}
	
	\begin{figure*}
		\centering
		\includegraphics[width=18cm]{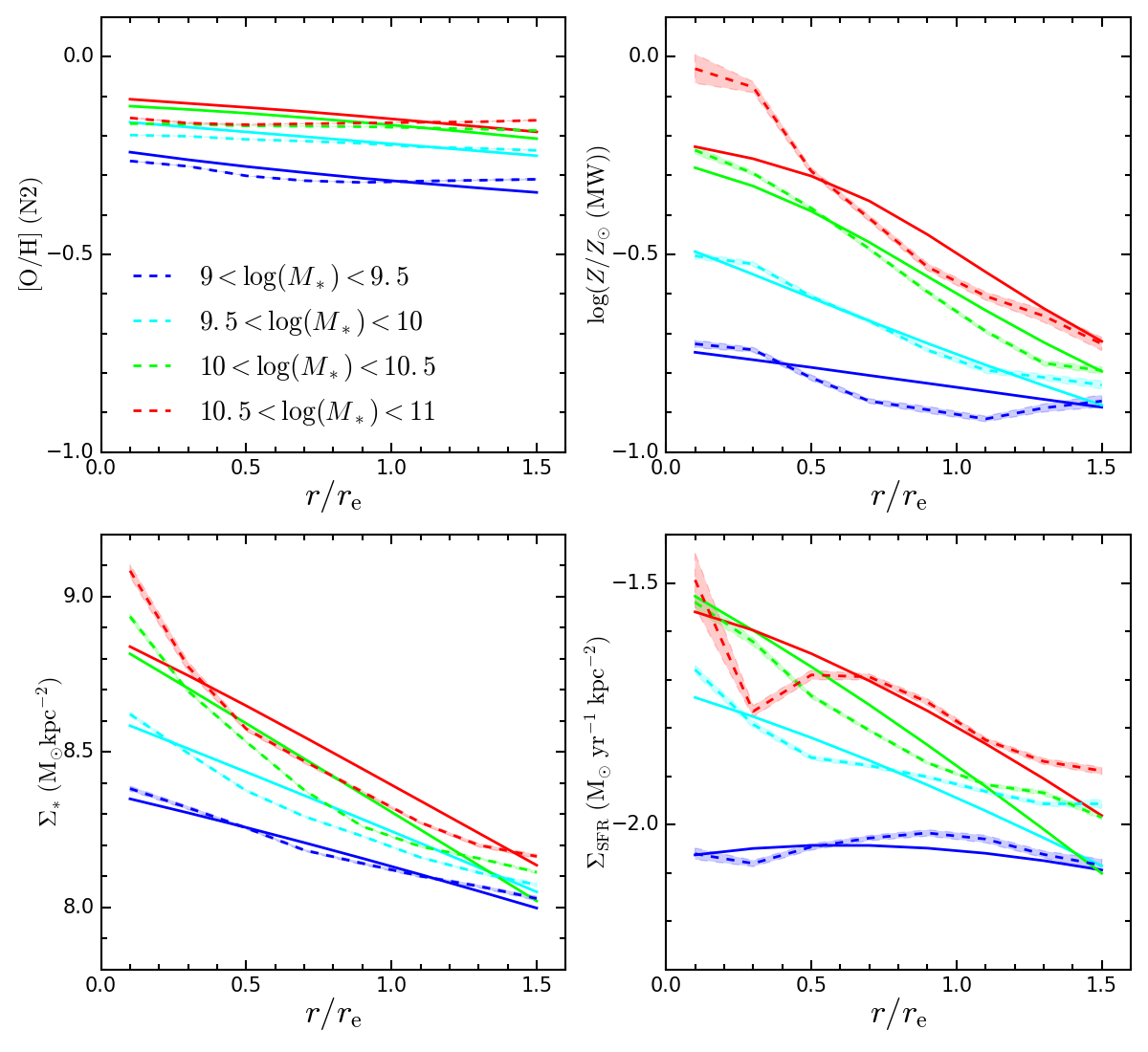}
		\caption{Similar figure to Figure A1 but showing the comparison between observations and the model with variable IMF slope at low mass end.
		}
		\label{figureA2}
	\end{figure*}
	
	\begin{table*}
		\caption{Parameter value of the time-dependent IMF model in Figure A2.}
		\label{TableA2}
		\centering
		\begin{tabular}{l c c c c c c c c c c}
			\hline\hline
			mass bin & $A_{\rm ks}$ & $A_{\rm inf}$ & $\tau_{\rm inf}$ & $f_{\rm out}$ & $t_{\rm out,e}$ & $\alpha1$ &$\alpha1_{\rm i}$ & $\alpha2$ & $\alpha2_{\rm i}$ \\ 
			log(${\rm M_{\odot}}$) & - & ${\rm M}_{\odot}{\rm yr^{-1}}$ & Gyr & - & Gyr & - & - & - & - \\
			\hline
			$[9,9.5]$ & 0.5 & 3$\frac{r}{r_{\rm e}}e^{-0.8\frac{r}{r_{\rm e}}}$ & 3+1.$\frac{r}{r_{\rm e}}$ & 0.78 & 13.7 & 1.3 & 2.6+0.3$\frac{r}{r_{\rm e}}$ & 2.3 & 2.3 \\
			$[9.5,10]$ & 1 & 3$\frac{r}{r_{\rm e}}e^{-1\frac{r}{r_{\rm e}}}$ & 5.5+1.5$\frac{r}{r_{\rm e}}$ & 0.65 & 13.7 & 1.3 & 2.6+0.9$\frac{r}{r_{\rm e}}$ & 2.3 & 2.3 \\
			$[10,10.5]$ & 1 & 10$\frac{r}{r_{\rm e}}e^{-1.4\frac{r}{r_{\rm e}}}$ & 5+1.6$\frac{r}{r_{\rm e}}$ & 0.64 & 13.7 & 1.3 & 1.7+1.5$\frac{r}{r_{\rm e}}$ & 2.3 & 2.3 \\
			$[10.5,11]$ & 1 & 26$\frac{r}{r_{\rm e}}e^{-1.3\frac{r}{r_{\rm e}}}$ & 5+2.1$\frac{r}{r_{\rm e}}$ & 0.64 & 13.7 & 1.3 & 1.2+1.8$\frac{r}{r_{\rm e}}$ & 2.3 & 2.3 \\
			\hline
		\end{tabular}\\  	
	\end{table*}
	\section{Variable high-mass IMF slope model}
	we show another viable IMF model with variable IMF slope at high mass end that reproduces the observed radial profile of gas metallicity (R23 method), stellar metallicity, mass surface density, and SFR surface density.
	
	\begin{figure*}
		\centering
		\includegraphics[width=18cm]{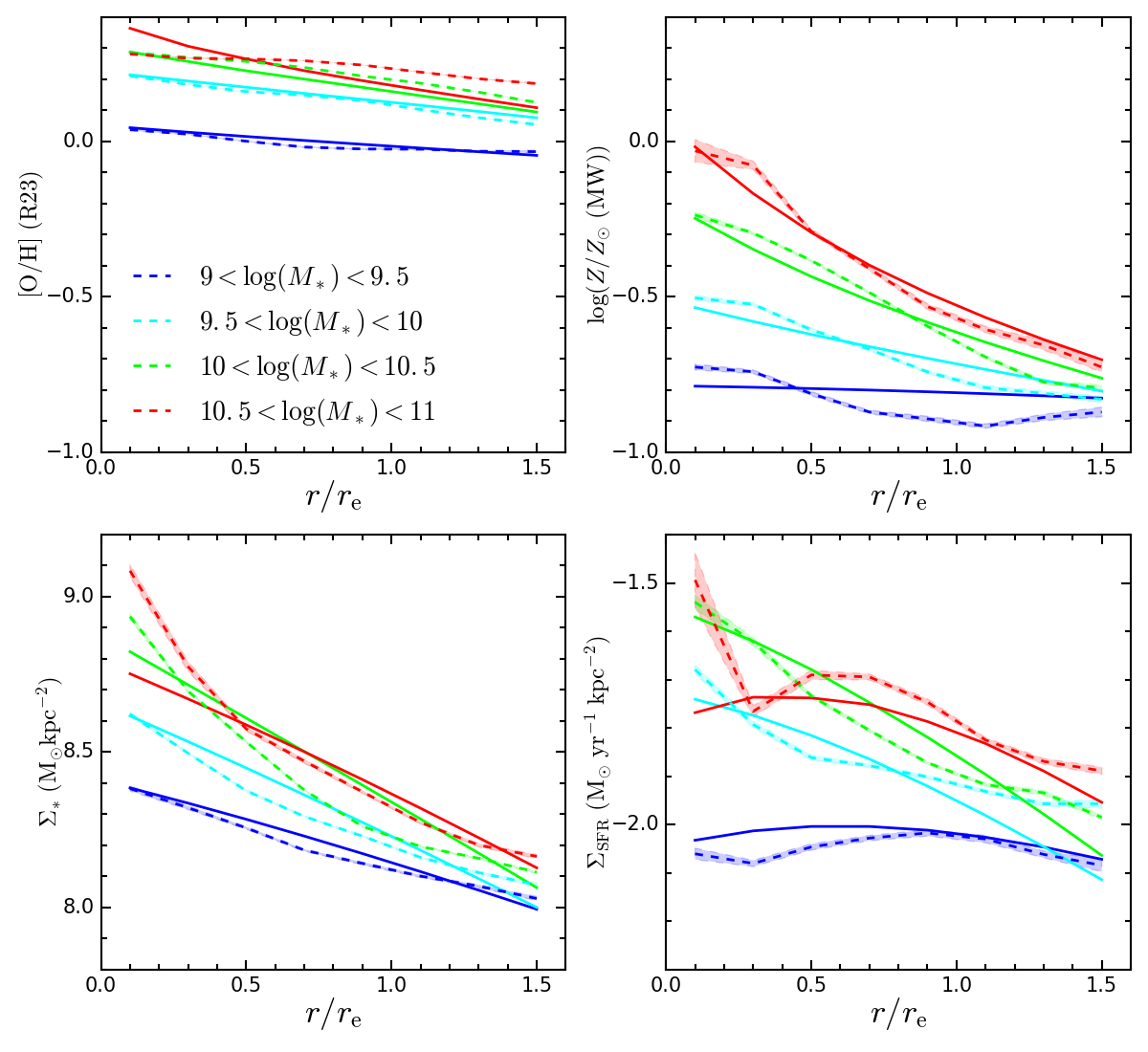}
		\caption{Similar figure to Figure 9 but showing the model with variable IMF slope at high mass end.
		}
		\label{figureB1}
	\end{figure*}
	
	\begin{table*}
		\caption{Parameter value of the time-dependent IMF model in Figure B1.}
		\label{TableB1}
		\centering
		\begin{tabular}{l c c c c c c c c c c}
			\hline\hline
			mass bin & $A_{\rm ks}$ & $A_{\rm inf}$ & $\tau_{\rm inf}$ & $f_{\rm out}$ & $t_{\rm out,e}$ & $\alpha1$ &$\alpha1_{\rm i}$ & $\alpha2$&$\alpha2_{\rm i}$ \\ 
			log(${\rm M_{\odot}}$) & - & ${\rm M}_{\odot}{\rm yr^{-1}}$ & Gyr & - & Gyr & - & - & - & - \\
			\hline
			$[9,9.5]$ & 1 & 3.$\frac{r}{r_{\rm e}}e^{-0.9\frac{r}{r_{\rm e}}}$ & 3.1+1.2$\frac{r}{r_{\rm e}}$ & 0.3 & 13.7 & 1.3 & 1.3 & 2.3 & 3.6+0.2$\frac{r}{r_{\rm e}}$ \\
			$[9.5,10]$ & 2 & 3.5$\frac{r}{r_{\rm e}}e^{-1.2\frac{r}{r_{\rm e}}}$ & 4.6+1.8$\frac{r}{r_{\rm e}}$ & 0 & 13.7 & 1.3 & 1.3 & 2.3 & 3.6+0.7$\frac{r}{r_{\rm e}}$ \\
			$[10,10.5]$ & 2 & 11$\frac{r}{r_{\rm e}}e^{-1.4\frac{r}{r_{\rm e}}}$ & 4.4+1.8$\frac{r}{r_{\rm e}}$ & 0 & 13.7 & 1.3 & 1.3 & 2.3 & 3+1$\frac{r}{r_{\rm e}}$ \\
			$[10.5,11]$ & 2 & 24$\frac{r}{r_{\rm e}}e^{-1.3\frac{r}{r_{\rm e}}}$ & 3.5+3$\frac{r}{r_{\rm e}}$ & 0 & 13.7 & 1.3 & 1.3 & 2.28 & 2.6+1.2$\frac{r}{r_{\rm e}}$ \\
			\hline
		\end{tabular}\\  	
	\end{table*}
	
\end{document}